\documentclass[journal,draftcls,onecolumn,12pt,twoside]{IEEEtranTCOM}
\normalsize

\usepackage{cite}

\ifCLASSINFOpdf
\else
\fi
\usepackage[pdftex]{graphicx}

\usepackage[cmex10]{amsmath}
\usepackage[cmintegrals]{newtxmath}
\usepackage{mathtools}
\usepackage{algorithm}
\usepackage{algpseudocode}
\usepackage{algorithmicx}
\usepackage[tight,footnotesize]{subfigure}

\usepackage{color}
\usepackage{soul}

\begin{document}
	\newcommand{\argmax}{\mathop{\rm arg~max}\limits}

	%
	\title{Repeat-Accumulate Signal Codes}
	%
	%
	%

	\author{Manato Takai,~\IEEEmembership{Student Member,~IEEE}
		and~Koji Ishibashi,~\IEEEmembership{Member,~IEEE}
		\thanks{M. Takai and K. Ishibashi are with the Advanced Wireless \&
			Communication Research Center (AWCC), The University of
			Electro-Communications, 1-5-1 Chofugaoka, Chofu-shi,
			Tokyo 182-8585, Japan e-mail: \{manato, koji\}@ieee.org}}

	%
	%

	\markboth{IEEE Transactions on Communications}%
	{Submitted paper}
	%



	\maketitle

	\begin{abstract}
		State-constrained
		signal codes directly encode modulation signals using signal processing
		filters, the coefficients of which are constrained over
		the rings of formal power series. Although the performance of signal codes is defined by these signal filters, optimal filters must be found by brute-force search in
		terms of symbol error rate because the
		asymptotic behavior with different filters has not been
		investigated. Moreover, computational complexity of the conventional BCJR used in the decoder increases
		exponentially as the number of output constellations increase. We hence propose a new class of state-constrained signal codes called \emph{repeat-accumulate signal codes} (RASCs). To analyze the asymptotic behavior of these codes, we employ Monte Carlo density evolution (MC-DE). As a result, the optimum filters can be efficiently found for given
		parameters of the encoder.
		We also introduce a low-complexity decoding algorithm for RASCs called the extended min-sum (EMS) decoder. The MC-DE analysis shows that the difference between noise thresholds of RASC and the Shannon limit is within 0.8 dB.
		Simulation results moreover show that the
		EMS decoder can reduce the computational complexity to less than 25\%
		of that of conventional decoder without degrading the performance by more than 1\,dB.
	\end{abstract}

	%
	\IEEEpeerreviewmaketitle

	\section{Introduction}
	\label{sec:intro}
	%
	%
	%
	%
	\IEEEPARstart{C}{apacity} approaching codes such as low-density
	parity check (LDPC) codes and turbo codes provide significant
	coding gains \cite{DE_1,Quantize_DE,EXIT}.
	Theoretical analysis shows that their noise thresholds, which are
	the maximum decodable noise variances, are close to the Shannon
	limit. Moreover, simulation results show that these codes exhibit
	high coding gain over an additive white Gaussian noise (AWGN) channel when
	binary phase shift keying is used \cite{LDPC_DE2,Turbo}.
	However, in practice, it is difficult to approach the Shannon capacity
	because the correlation among unreliable coded bits degrades the decoding
	performance when high-order modulation is used.
	Coded modulation is an effective technique for enhancing the coding gain with high-order modulation because this technique enables the Hamming distances of the linear codes and the Euclidean distances of the modulation to be designed in an integrated manner \cite{Ungerboeck, BICM}.
	Unfortunately, the computational complexity required to design these two distances such that a high transmission rate can be achieved is impractical because of the enormous number of constellation points involved.

	Lattice codes are structured codes for AWGN channels and
	have gained a great deal of attention since Urbanke and Rimoldi proved that lattice codes can achieve any rate less than the Shannon capacity with
	arbitrarily small probability \mbox{\cite{LatticeAWGNCapacity}}.
	The advantage of lattice codes is that
	linear codes over a Hamming space can easily be transformed into
	Euclidean space by an algebraic construction \cite{Lattice}.
	Construction A lattices with lattice shaping have been proposed for average power constrained channels \mbox{\cite{QC-LDPC-lattice, Leech_ConstA, IRA_lattice}}.
	For instance, the gap between the symbol error rate (SER) performance of irregular repeat-accumulate lattices proposed in \cite{IRA_lattice} and the Shannon limit is 0.48 dB. (In this case, the Shannon limit can be interpreted as the maximum decodable noise variance with an average power constraint on the input signals.) Unfortunately, although the lattices based on Construction A can be decoded by a $q$-ary decoder similar to non-binary codes, $q$ must be prime. Hence, Construction A is not suitable for typical binary-based communications systems. To overcome this limitation, Construction D was proposed in \mbox{\cite{ConstD}}. This construction is based on a nested family of binary codes and is suitable for common communications systems. For instance, \emph{turbo lattices} achieve a performance that is within approximately 1.25\,dB of the Poltyrev limit, which is derived as the unconstrained capacity of AWGN and reveals the maximum decodable noise variance \emph{without} any power constraints. However, the decoder of Construction D consists of not only multiple binary decoders but also the corresponding binary encoders, interference cancelers, and modulo operators. This inherent nested architecture obviously leads to high complexity \mbox{\cite{Turbo_lattice_dec, LDPC_lattice}}.

	Recently, signal codes have been proposed as an alternative lattice-coded modulation technique.
	The entire encoding process consists of a simple
	signal convolution via infinite impulse response (IIR) and/or finite
	impulse response (FIR) filters.
	In signal codes, the filter coefficients are the most important
	parameters for coding gain because these coefficients define the Euclidean
	distances among generated codewords.
	In \cite{CLC}, it was shown that
	the absolute values of filter coefficients must be close to one in order to
	obtain high coding gain. This implies that the transmission
	power exponentially increases with codeword length, and shaping
	techniques such as Tomlinson-Harashima precoding (THP) are necessary to
	meet an average power constraint in practice \cite{THP1, Harashima}.
	Although THP can tailor the
	transmission power, the number of possible constellation points becomes
	infinite. This fact results in an exponential decoding complexity, and
	well-known powerful decoding algorithms such as the BCJR algorithm cannot
	be utilized \cite{SPA}.
	Therefore, in \cite{CLC}, a list decoder is used, and the
	decoding performance is within 2\,dB
	of the Shannon limit with a frame error rate (FER) of
	$10^{-3}$. However, list decoding is sub-optimal and is inferior to
	the BCJR algorithm.

	Mitran and Ochiai proposed \emph{state-constrained} signal codes
	called {\em turbo signal codes} to overcome the explosion of
	decoding complexity \cite{TSC}. In \emph{state-constrained} signal
	codes, shaping coefficients are chosen over a coset leader of the rings of formal power series.
	Because of the group properties, the output signals
	are also constrained over the coset leader, and the number of possible
	constellation points becomes finite. This additional constraint enables
	the decoder to use the BCJR algorithm, and the resulting performance is within
	0.8\,dB of the Shannon limit with an SER of
	$10^{-4}$.
	Although the decoding complexity has been reduced by the use of rings,
	turbo signal codes still suffer from a remarkable increase in the number of possible states of
	filters when higher-order modulation is used as input because the number
	of states increases exponentially as the order of the modulation
	increases. This prevents the decoder from using the BCJR
	algorithm. Moreover, the fundamental code properties
	of signal codes, such as the noise threshold, have not yet been
	investigated.
	Therefore, to find the optimum filter coefficients in terms of coding gain, a brute-force search is unavoidable \cite{TSC}.

	In this paper, to address the inherent problems of conventional turbo signal codes, we propose novel state-constrained signal codes called
	\emph{repeat-accumulate signal codes} (RASCs). RASCs are based on
	repeat-accumulate (RA) codes \cite{RA}.
	The proposed RASCs are composed of a repeater,
	an interleaver, and a one-tap IIR filter, i.e., an accumulator. Here, the codewords are restricted not only by the rings of formal power series
	but also by parity check constraints. Although the parity check constraints
	require density evolution in order to analyze the noise threshold
	of the RASCs, regular density evolution cannot be used because tracking true densities is
	impractically complex. Hence, we employ MC-DE \cite{Matteo_phd} to
	determine the asymptotic behavior of the RASCs. The conditions for the
	optimum filter are also determined using MC-DE. Moreover, we introduce
	a low-complexity decoding algorithm based on the EMS algorithm.
	The proposed algorithm is modified to decode the codes over the coset leader of the rings of formal power series.
	The simulation results reveal that the modified EMS can reduce the complexity to a quarter of that of the BCJR algorithm without significant performance loss.
	In summary, the contributions of this study are as follows:
	\begin{itemize}
		\item A new class of signal codes, called RASCs, are proposed.
		\item The noise thresholds for several important RASC parameters such as
		the filter, ring, column weight, and
		input constellation size, are determined using MC-DE.  By carefully choosing the parameter values,
		the Shannon capacity can be approached to within approximately 0.8\,dB. Moreover, numerical results demonstrate that the performance RASC approaches within 1.5\,dB of the Shannon limit with an information length of 1,000, supporting this analysis.
		\item Our analysis via MC-DE leads us to introduce a new input constraint for increasing the transmission rate. In conventional turbo signal codes, the input
		signals are constrained by quadrature amplitude modulation (QAM),
		which is a subset of the coset leader of the rings of formal power series because
		providing redundancy in terms of constellation size can be
		used to improve the error performance
		\cite{TSC}. However, we show
		that the noise thresholds when the constellation of the input signals are identical to the coset leader are slightly better than those when the input
		signals are constrained by QAM.
		\item We introduce a modified EMS algorithm for the proposed
		codes to reduce the decoding complexity \cite{EMS}.
		The key concept of the modified EMS algorithm is to extract only the highly reliable elements of the
		message vector in order to reduce the size of the exchanged message vector.
		As a result, the decoding complexity dramatically decreases because it
		does not depend on the number of output constellation points but rather
		on the size of the truncated message vectors.
		Simulation results show that the performance loss caused by message truncation does not exceed 1\,dB whereas its computational complexity is less than a quarter of that of the BCJR algorithm or FFT-BP algorithm.
	\end{itemize}

	The remainder of this paper is organized as follows. The basics of the signal codes such as
	the system model, channel model, and definition of the rings of formal power series are
	explained in Section \ref{sec:system}. In Section \ref{sec:RASC}, we
	propose the encoding and decoding structures of RASCs. Section
	\ref{sec:Threshold} presents the optimum filters and their noise
	thresholds as calculated by MC-DE.  In Section \ref{sec:Results}, the numerical
	results of RASCs obtained using the FFT-BP and EMS algorithms are presented,
	and the decoding performances of RASCs and turbo signal
	codes are compared. Finally, Section \ref{sec:Conclusion} concludes the paper.

	\subsection{Notation}
	Throughout the paper, variables are expressed as follows:
	a vector is indicated by boldface in lower case, a matrix is represented by boldface in upper case unless otherwise specified, $\mathbb{Z}$ represents a set of integers, and $\mathbb{C}$ indicates a set of complex numbers. Let $(\cdot)^\mathrm{T}$ denote the transpose of a matrix or a vector and $\boldsymbol{E}\left[\cdot \right]$ denote the mean of a random variable.

	\section{Basics of Signal Codes}
	\label{sec:system}
	\subsection{Original signal codes}
\begin{figure}[tb]
		\begin{center}
			\includegraphics[width=\hsize]{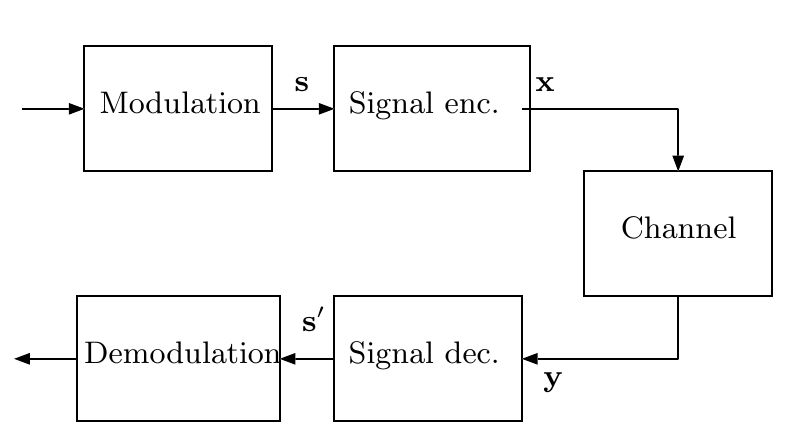}
			\caption{System model of signal codes}
			\label{system}
		\end{center}
	\end{figure}

	Figure \ref{system} shows a system model of signal codes.
	Input complex signals $\boldsymbol{s}=\left(s_1,s_2,\cdots,s_{N_s}\right)^T\in
	\mathbb{C}^{N_s\times 1}$ are directly encoded by a signal encoder, where $N_s$
	is the number of input signals.
	For the input signals, $L^2$-QAM \footnote{The defined $L^2$-QAM is one generalized form of QAM constellations. This is a different constellation from \emph{square}-QAM, which is the most popular one, defined as $\{-L+1,\cdots,-1,1,\cdots,L-1\}$ for both the real and imaginary parts of coefficients.
	} is defined as follows:
	\begin{equation}
		\mathbb{Z}_L[j]=\left\{a+jb:a,b\in\mathbb{Z}/L\mathbb{Z}\right\},
	\end{equation}
	where $j=\sqrt{-1}$.
	The encoder maps the input signals to coded signals $\boldsymbol{x}=\left(x_1,x_2,\cdots,x_{N_c}\right)^T\in \mathbb{C}^{N_c\times 1}$ via
	a generator matrix $\boldsymbol{G}\in \mathbb{C}^{N_c\times N_s}$,
	which has the following Toeplitz form:
	\begin{equation}
		\boldsymbol{G}=\left(\begin{array}{ccccccc}
			1      & 0      & \cdots&\cdots &0 & 0      & \\
			g_1    & 1      & \ddots&\ddots &\vdots & \vdots & \\
			\vdots & g_1    & \ddots&\ddots &\vdots & \vdots & \\
			g_p    & \vdots & \ddots&\ddots &1 & \vdots & \\
			0      & g_p    & \ddots&\ddots &g_1 & 1 & \\
			\vdots & 0      & \ddots&\ddots &\vdots & g_1 & \\
			\vdots & \vdots & \ddots&\ddots &g_p    & \vdots & \\
			0      & 0      & \cdots&\cdots &0      & g_p    & \\

		\end{array}\right),
	\end{equation}
	where $1,g_1,g_2,\cdots,g_p$ correspond to the impulse response coefficients of
	an FIR filter. Thus, the resulting codes are said to be {\em convolutional
		lattice codes}.
	To meet the average power constraint, a shaping operation
	should be performed in the encoding process. A shaping vector
	$\boldsymbol{b}\in \mathbb{C}^{N_c\times 1}$ restricts the amplitude of encoded
	signals $x_i$, $i\in \{1,\cdots,N_c\}$, into a desired shaping
	region. For the original signal codes, THP is assumed to be the shaping
	operation. The aim of the THP operation is to restrict the real and imaginary parts of the output
	constellation to $[0,L)$. The shaping operation is hence
	\begin{equation}
		x_t = s_t + \sum_{k=1}^{N_{FF}} g_k\left(s_{t-k} + Lb_{t-k}\right) +Lb_t,
	\end{equation}
	where $N_{F}$ is the number of filter taps and the $b_t$ are chosen as follows:
	\begin{eqnarray}
		b_t &=& -L\left \lfloor \frac{1}{L}\left(s_t +
		\sum_{k=1}^{N_F}g_k\left(s_{t-k} +
		Lb_{t-k}\right)\right)
		\right\rfloor.
	\end{eqnarray}
	Thus, the $t$th received signal $y_t$ over the AWGN channel can be written as
	\begin{equation}
		y_t = x_t + z_t,
	\end{equation}
	where $z_t$ is an independent and identically distributed circularly-symmetric complex Gaussian random variable with mean zero and variance $\sigma^2$.
	\subsection{Turbo signal codes}
	Turbo signal codes introduce the constraint based on
	the ring of a formal power series to control the number of output
	constellation points.
	The ring of formal power series considered in this study is given by
		\begin{equation}
			\mathbb{Z}\left[e^{j\frac{\pi}{2N_{bv}}}\right]=\left\{\sum_{i=0}^{N_{bv}-1}\left(v_i^I+jv_i^Q\right)e^{j\frac{i\pi}{2N_{bv}}}\colon
			v^I_i,v^Q_i \in \mathbb{Z}
			\right\},
			\label{eq:fps_ring}
		\end{equation}
		where $N_{bv}$ is an integer indicating the number of possible phase shifts $e^{j\frac{\pi}{2N_{bv}}}$ that works as the indeterminate in the defined formal power series.
		In this ring, addition and multiplication for $A=\sum_{i=0}^{N_{bv}-1} (a_i^I+ja_i^Q)e^{\frac{ji\pi}{2N_{bv}}}$ and $B=\sum_{i=0}^{N_{bv}-1} (b_i^I+jb_i^Q)e^{\frac{ji\pi}{2N_{bv}}}$ can be defined as
		$A+B=\sum_{i=0}^{N_{bv}-1} \left((a_i^I+ja_i^Q)+(b_i^I+jb_i^Q)\right)e^{\frac{ji\pi}{2N_{bv}}}$ and
		$A\times B = \sum_{i=0}^{N_{bv}-1} \sum_{k=0}^{N_{bv}-1}(a_i^I+ja_i^Q)(b_{k}^I+jb_{k}^Q)e^{\frac{j(i+k)\pi}{2N_{bv}}}$, respectively.
		In addition, the set of formal power series modulo $L$ that is a coset leader of $\mathbb{Z}\left[e^{j\frac{\pi}{2N_{bv}}}\right]$ can be defined by
		\begin{eqnarray}
			&&C\left(L,N_{bv}\right)\nonumber \\
			&&=\mathbb{Z}\left[e^{j\frac{\pi}{2N_{bv}}}\right]/L\mathbb{Z}\left[e^{j\frac{\pi}{2N_{bv}}}\right]
			\nonumber \\
			&&=\left\{\sum_{i=0}^{N_{bv}-1}\left(v_i^I+jv_i^Q\right)e^{j\frac{i\pi}{2N_{bv}}} \colon v^I_i,v^Q_i \in \left\{0,1,\cdots,L-1\right\}
			\right\}. \nonumber \\
			\label{eq:fps}
		\end{eqnarray}
		Although $C\left(L,N_{bv}\right)$ is not a ring, the $t$th coded signal of turbo signal codes can be restricted to $C\left(L,N_{bv}\right)$ by shaping coefficient $b_t\in C\left(L,N_{bv}\right)$, which is based on THP and given by
		\begin{eqnarray}
			x_t &=& \sum_{k=1}^{N_F} f_{k}u_{t-k}+f_0\left(s_t+\sum_{k=1}^{N_F}
			g_{k}u_{t-k}+Lb_t\right), \\
			u_t &=& s_t + \sum_{k=1}^{N_F}g_ku_{t-k} + Lb_t,
			\label{eq:shap}
	\end{eqnarray}
	where $g_{k}$ and $f_{k}$ respectively indicate the feedback and feedforward
	filter coefficients constrained in $\mathcal{C}\left(L,N_{bv}\right)$ and $u_{t-k}$ represents the signal held in the $k$-th memory of the encoder.
	The cardinality of $\mathcal{C}\left(L,N_{bv}\right)$ is obtained by
	\begin{equation}
		\left|\mathcal{C}\left(L,N_{bv}\right)\right| = L^{2N_{bv}}.
	\end{equation}

	We define the index of the feedback filters FB for concise notation.
	From (\mbox{\ref{eq:fps}}), each filter consists of $2N_{bv}$ integer coefficients,
	$v_i^I$ and $v_i^Q$ for $i=0,1,\cdots,N_{bv}-1$.
	Therefore, if the index of filters is composed of $v_i^I$ and $v_i^Q$,
	this index can identify the set of the filter taps. For this reason, the index FB is defined as
	\begin{equation}
		\text{FB} = \sum_{j=1}^{N_F}\sum_{i=0}^{N_{bv}-1}\left(v_i^I\times L^{2i+(j-1)*2N_{bv}}+v_i^Q\times L^{2i+1+(j-1)*2N_{bv}}\right).
	\end{equation}

	For instance, when the parameters are defined as $L=2$, $N_{bv}=2$,
	$N_F=1$, and $g_1=(1+j)+(0+j)e^{j\pi/4}$, the index of the filter is
	$\text{FB}=1\times 2^0+1\times 2^1 + 0\times 2^2 + 1\times 2^3=11$.

	The index of the feedforward filters FF is expressed in the same  manner as
	FB. Hence,
	\begin{equation}
		\text{FF} = \sum_{j=0}^{N_F}\sum_{i=0}^{N_{bv}-1}\left(v_i^I\times L^{2i+j*2N_{bv}}+v_i^Q\times L^{2i+1+j*2N_{bv}}\right).
	\end{equation}

	\section{Repeat-Accumulate Signal Codes}
	\label{sec:RASC}
	At least three-tap filter coefficients
	are required to construct turbo signal codes, namely, two tap coefficients
	for the feedforward filters $f_0$ and $f_1$ and one tap coefficient for the feedback filter
	$g_1$. It is difficult to analyze the decoding
	performance and design the filter coefficients for turbo signal codes because the search space
	of the filter coefficients has a minimum of $L^{6N_{bv}}$ candidates.
	In contrast, the proposed RASCs reduce the search space to
	$L^{2N_{bv}}$ because our encoder has only one-tap filter
	coefficient.

	\subsection{Encoding Structure}
	\label{sec:encoding}
	\subsubsection{Encoder}
\begin{figure}[t]
		\begin{center}
			\includegraphics[width=\hsize]{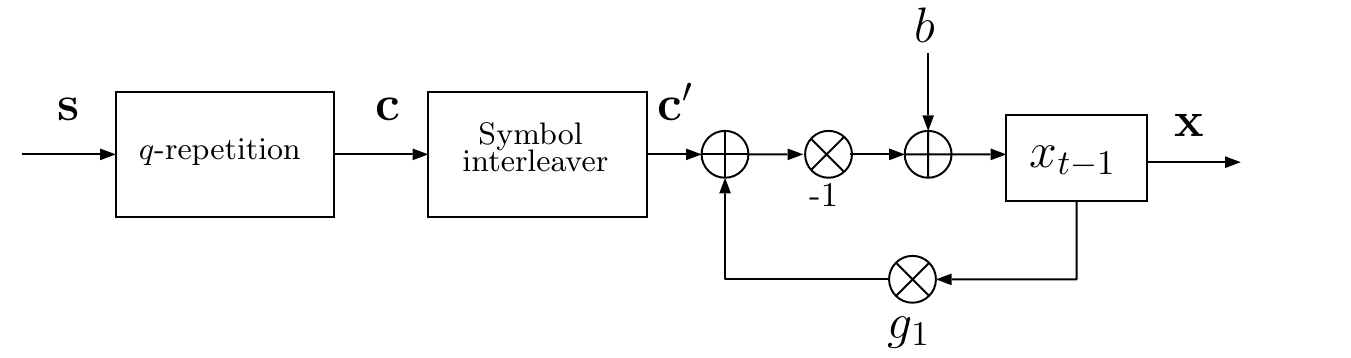}
			\caption{Encoder of the non-systematic RASC}
			\label{encoder}
		\end{center}
	\end{figure}

	The encoder of the RASC is shown in Fig. \ref{encoder}. In this paper, a non-systematic RASC is assumed so that only parity signals will be sent. The proposed RASC can be easily extended to systematic codes by sending information signals similar to binary RA codes \mbox{\cite{Johnson}}.
	The figure shows that the encoder consists of
	the same components that are found in RA codes.
	The output signal from the $q$-times repetition encoder is
	$\boldsymbol{c}=\left(c_1,c_2,\cdots,c_{qN_s}\right)^T$ and the output
	signal from the
	interleaver is $\boldsymbol{c'}=\left(c'_1,c'_2,\cdots,c'_{qN_s}\right)^T$, both of which
	are constrained over the same set as the input signals $\boldsymbol{s}$. For
	instance, when the input
	signals are $\mathbb{Z}_L[j]$, $\boldsymbol{c}$ and
	$\boldsymbol{c'}$ become
	$\left(\mathbb{Z}_L\left[j\right]\right)^{N_c}$, where $N_c = qN_s$
	because non-systematic RASC is assumed.
	The accumulator of the RASC
	is slightly different from that of the binary RA code because the accumulator consists of a filter
	and a shaping operation. The $t$th output signal from the encoder is given by
	\begin{equation}
		x_t = -\left(c'_t + g_1x_{t-1}\right)+Lb_t.
	\end{equation}
	The resulting codes satisfy the following parity check equation:
	\begin{equation}
		x_t + c'_t + g_1x_{t-1} = 0 \mod
		C\left(L,N_{bv}\right),
		\label{eq:pq}
	\end{equation}
	where the operation $\mod C\left(L,N_{bv}\right)$
	is calculated by a modulo-$L$ operation for each coefficient $v_i^I$ and $v_i^Q$.
	An RASC can be decoded by the sum-product algorithm because the parity check matrix can be defined based on (\mbox{\ref{eq:pq}}).
	In contrast, conventional turbo signal codes with $L=2$ and $N_{bv}=2$ do not pass the parity check, so turbo signal codes cannot be decoded by the sum-product decoder. This proof is presented in detail in Appendix \mbox{\ref{sec:pr_turbo}}. Note that to obtain higher coding gain, the following four parameters must be optimized:
	the number of repetitions $q$, number of phase shifts $N_{bv}$, size of input constellation $L$, and filter coefficient $g_1$.

	Note that throughout this paper, we assume that the RASC considered in this study has a code rate of less than half because of the non-systematic structure. Although puncturing or a combiner could increase the coding rate so that is similar to that of binary RA codes, a systematic structure would be necessary to use these techniques \mbox{\cite{Johnson}}.

	\subsubsection{Transmission Power}
	Encoded signals $\boldsymbol{x}$ are obtained
	over the set $C\left(L,N_{bv}\right)$, as
	described above.
	%
	However, this constellation requires unnecessarily high transmission power because its center is not at the origin of the complex plane.
	Therefore, to minimize the average transmission power, the constellation is shifted so that it is centered at the origin.
		The shifted constellation is defined as $\tilde{C}\left(L,N_{bv}\right)\in \mathbb{C}$. Figure \ref{const} illustrates several signal constellations of $\tilde{C}(L,N_{bv})$. Note that the receiver easily transforms $\tilde{C}\left(L,N_{bv}\right)$ into $C\left(L,N_{bv}\right)$. However, because of this transformation, the merged constellation points in $\tilde{C}\left(L,N_{bv}\right)$ are treated as the same element in $C\left(L,N_{bv}\right)$ \cite{TSC}.
	The resulting transmission power of $\tilde{C}(L,N_{bv})$ is given by \cite{TSC}
	\begin{equation}
		E\left[\left|\boldsymbol{x}\right|^2\right] = \frac{N_{bv}\left(L^2-1\right)}{6}.
	\end{equation}

	\subsubsection{Constellation of RASC}
	\label{sec:const}
\begin{figure*}[t]
		\begin{center}
			\subfigure[$\mathcal{\tilde{C}}(2,2)$]{%
				\includegraphics[width=0.2\hsize]{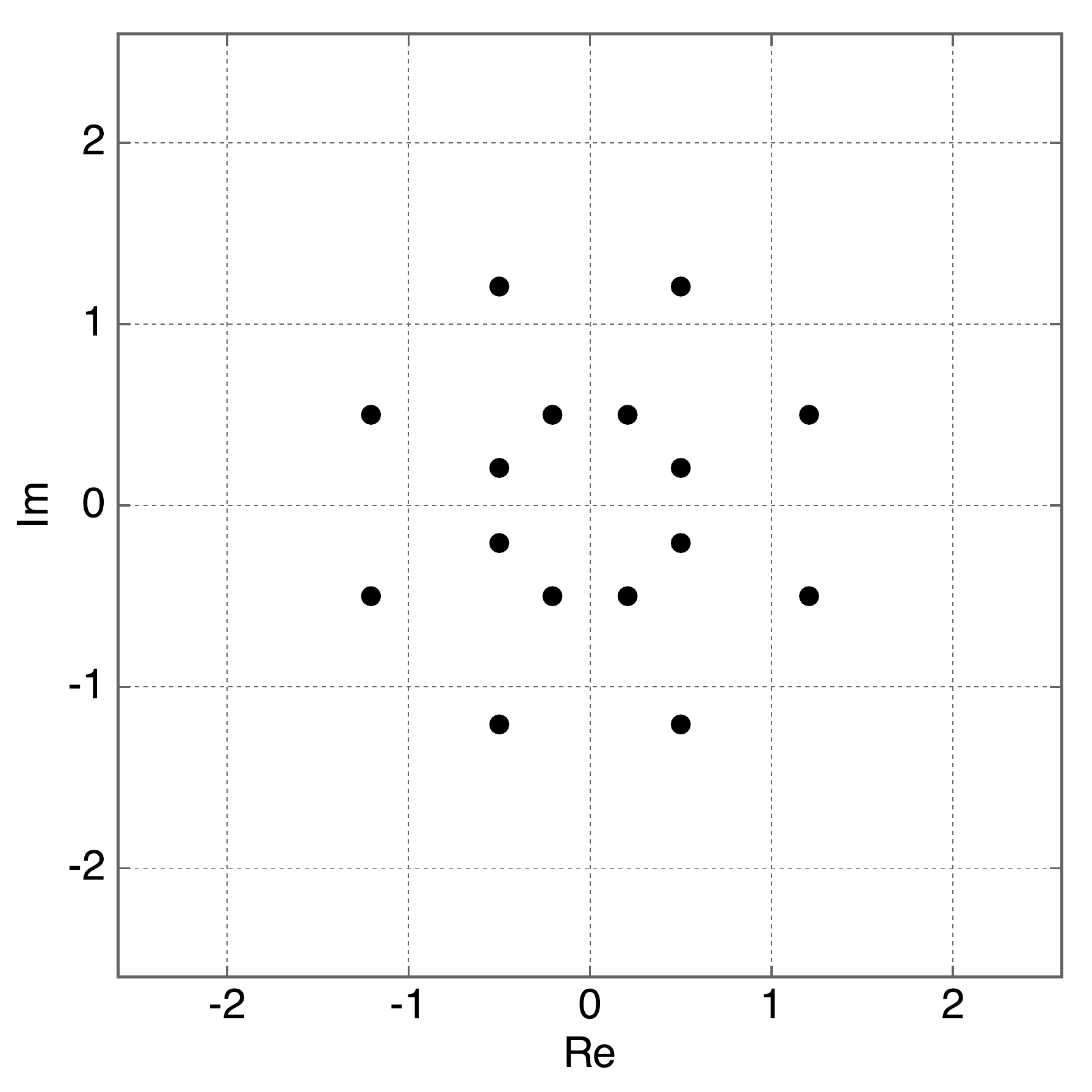}
				\label{const_L2N2}
			}
			\subfigure[$\mathcal{\tilde{C}}(2,3)$]{%
				\includegraphics[width=0.2\hsize]{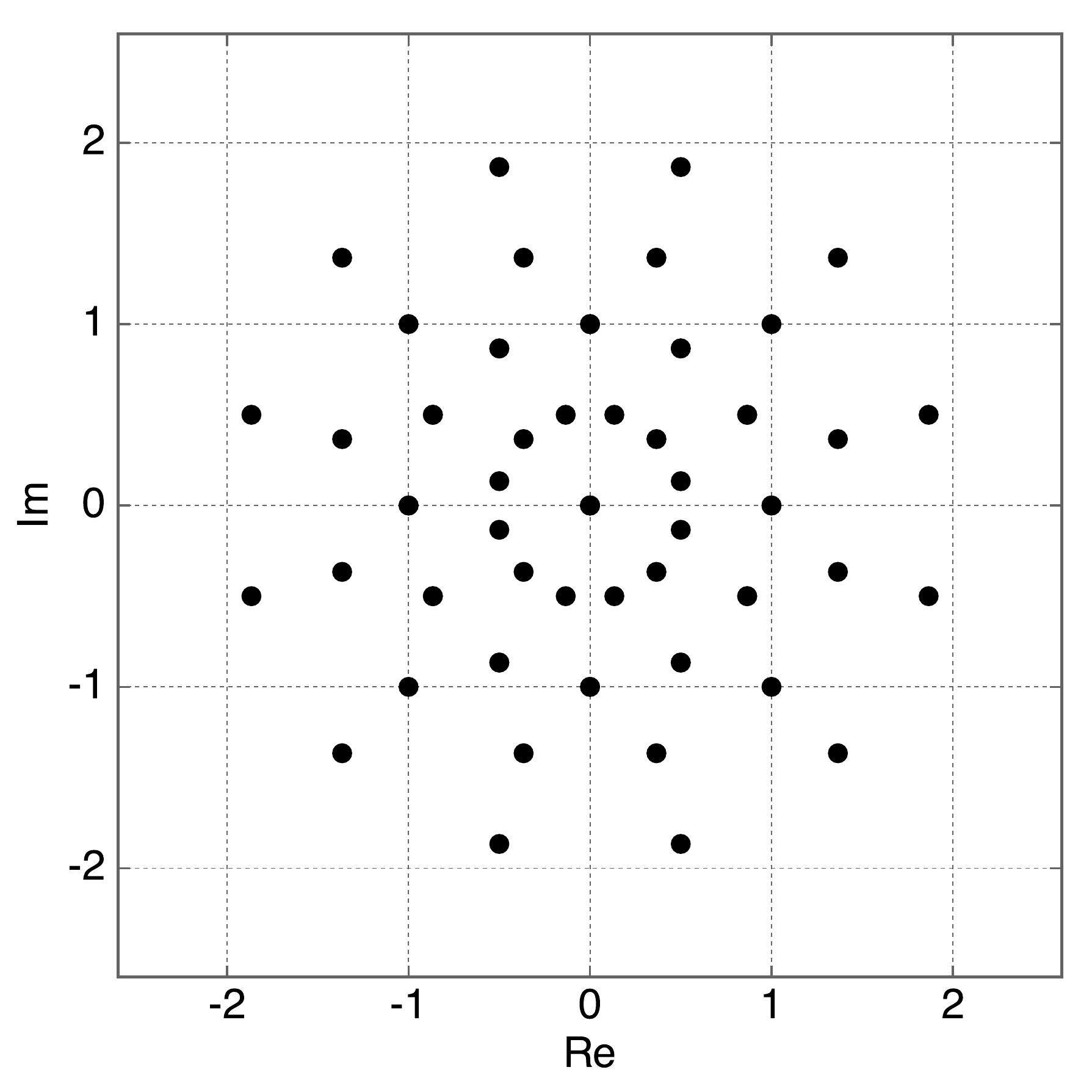}
				\label{const_L2N3}
			}
			\subfigure[$\mathcal{\tilde{C}}(2,4)$]{%
				\includegraphics[width=0.2\hsize]{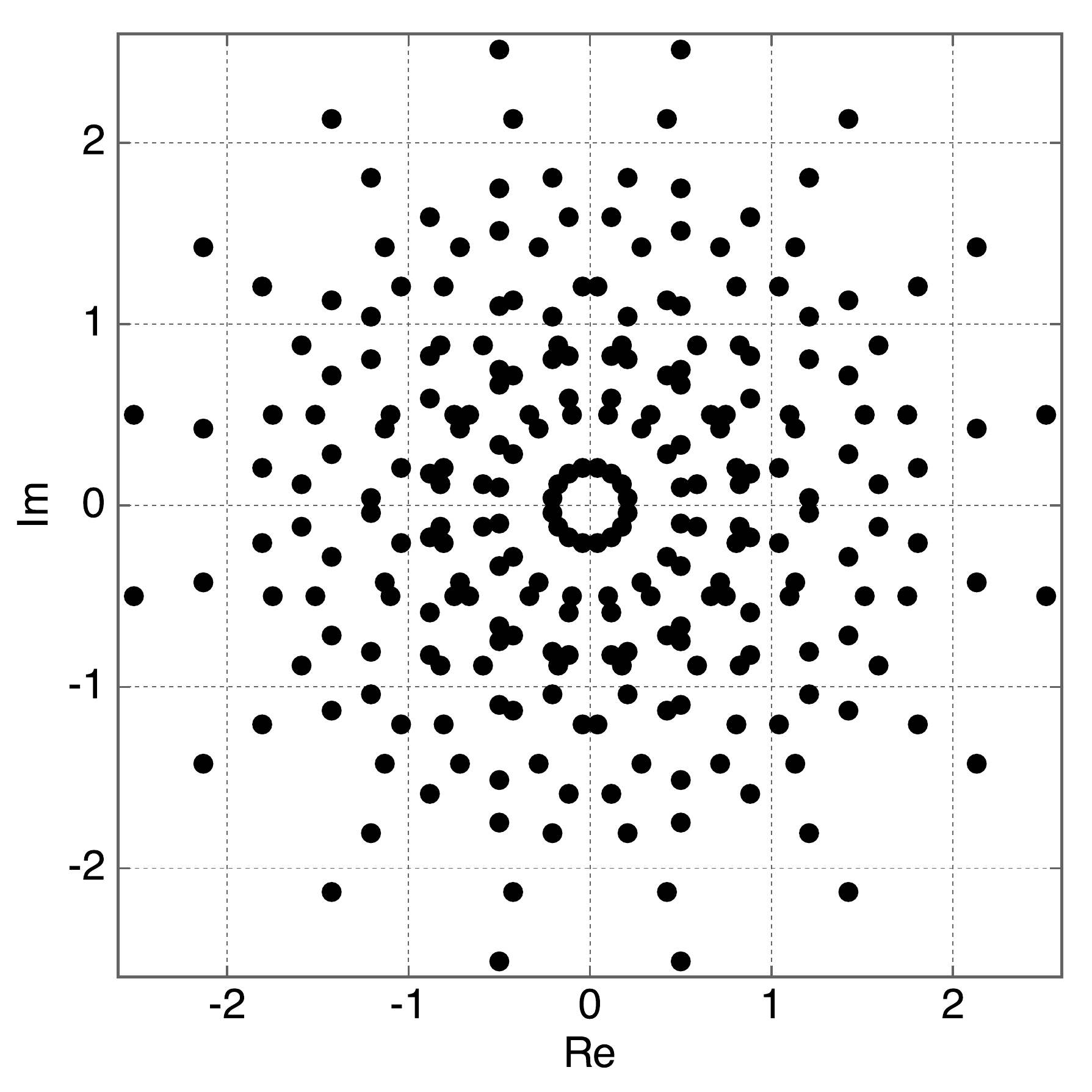}
				\label{const_L2N4}
			}
			\subfigure[$\mathcal{\tilde{C}}(3,2)$]{%
				\includegraphics[width=0.2\hsize]{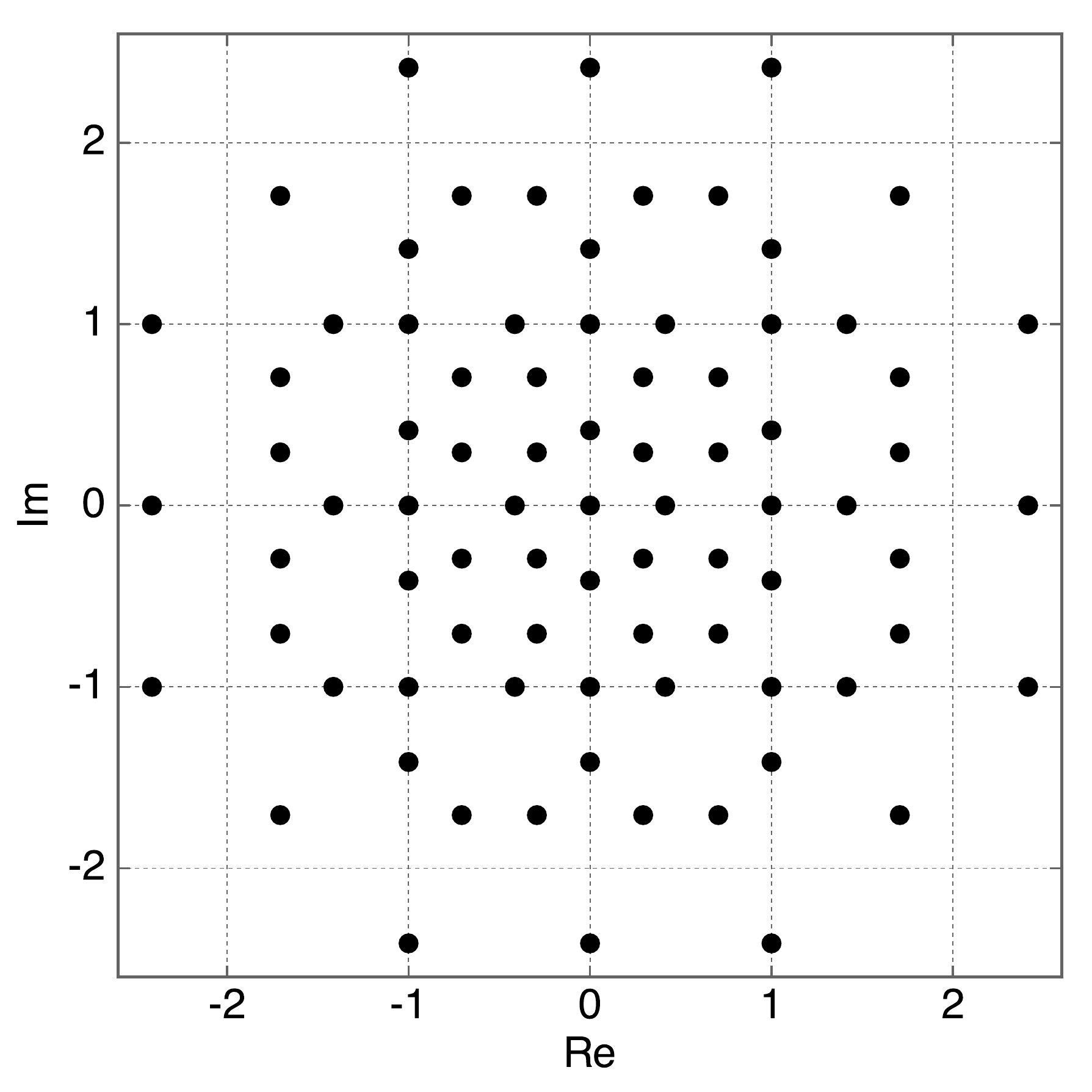}
				\label{const_L3N2}
			}
			\caption{Examples of all possible constellation points at the output of the RASC encoder with $\mathcal{\tilde{C}}(L,N_{bv})$.}
			\label{const}
		\end{center}
	\end{figure*}

	As shown in Fig. \ref{const}, the number of signal points
	appears to increase as $L$ or $N_{bv}$ increases.
	Several points of
	$C(L,N_{bv})$ are transformed into the same point of $\tilde{C}(L,N_{bv})\in \mathbb{C}$ when $N_{bv}>2$ because the calculation rule for set $C\left(L,N_{bv}\right)$ is different from that of the complex field.
	For instance, when $L=2$ and $N_{bv}=3$, two points $a=(0+j)+(0+0j)e^{j\pi/6}+(0+j0)e^{j2\pi/6}$ and
	$b=(0+j)+(1+0j)e^{j\pi/6}+(0+j)e^{j2\pi/6}$ cannot be further simplified and thus become different points in set $C(L,N_{bv})$.
	However, these points are the same point when $\tilde{C}(L,N_{bv})\in\mathbb{C}$,
	namely $a=b=(0+j)$.
	Therefore, in Figs. \mbox{\ref{const}}-\mbox{\subref{const_L2N3}} and \mbox{\ref{const}}-\mbox{\subref{const_L2N4}}, the number of
	constellation points is less than $L^{2N_{bv}}$.

\begin{figure*}[t]
		\begin{center}
			\subfigure[FB = 3]{%
				\includegraphics[width=0.2\hsize]{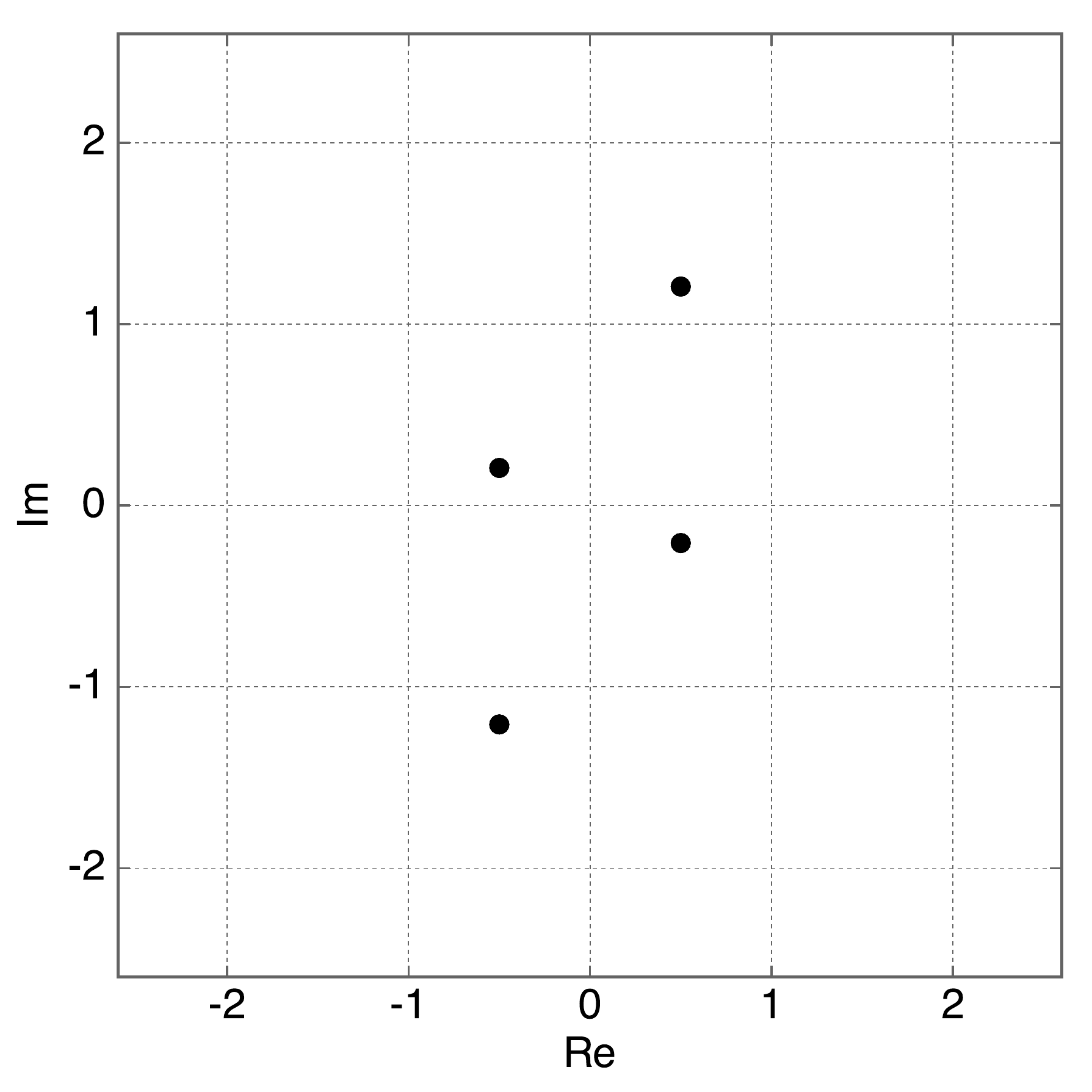}
				\label{const_FB3}
			}
			\subfigure[FB = 5]{%
				\includegraphics[width=0.2\hsize]{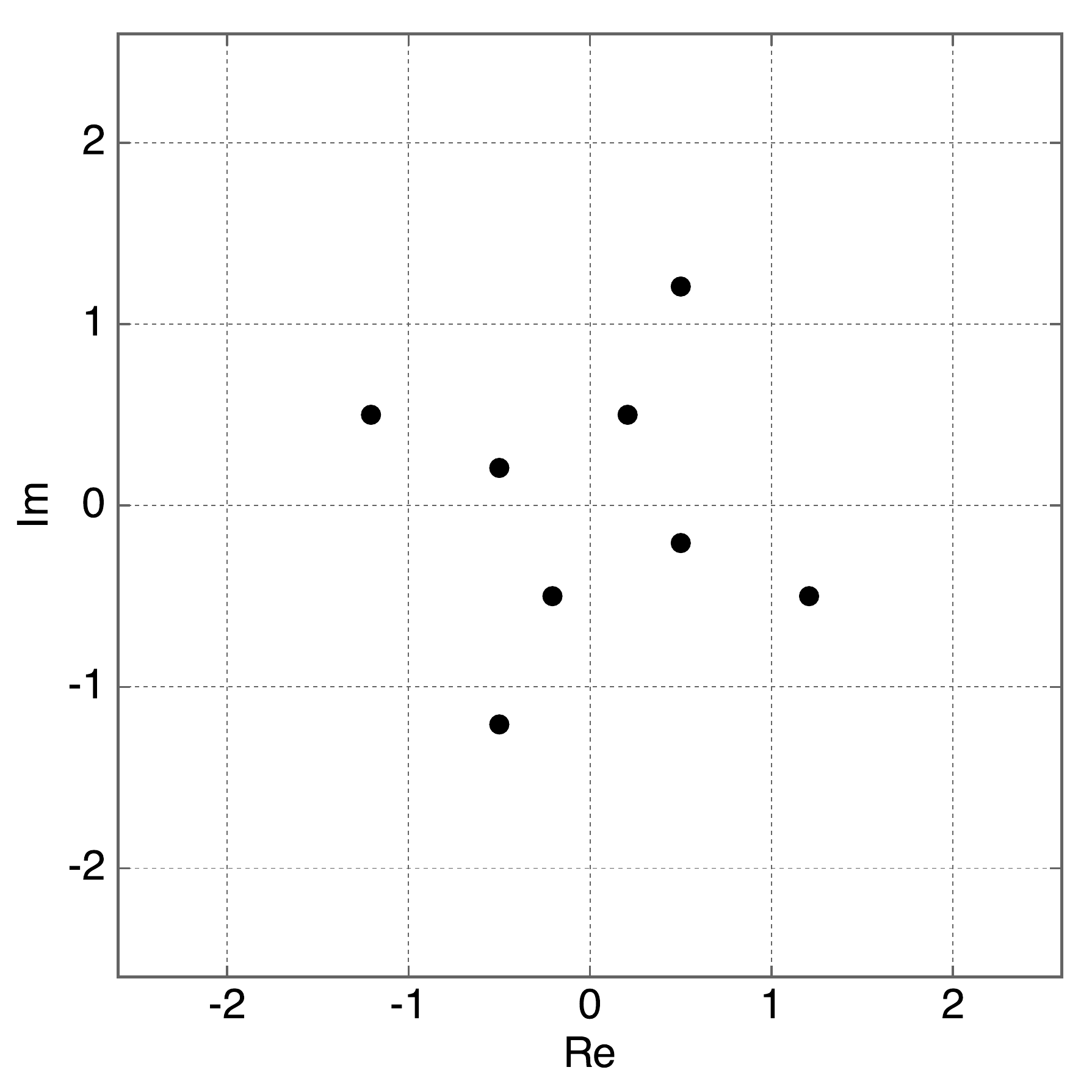}
				\label{const_FB5}
			}
			\subfigure[FB = 11]{%
				\includegraphics[width=0.2\hsize]{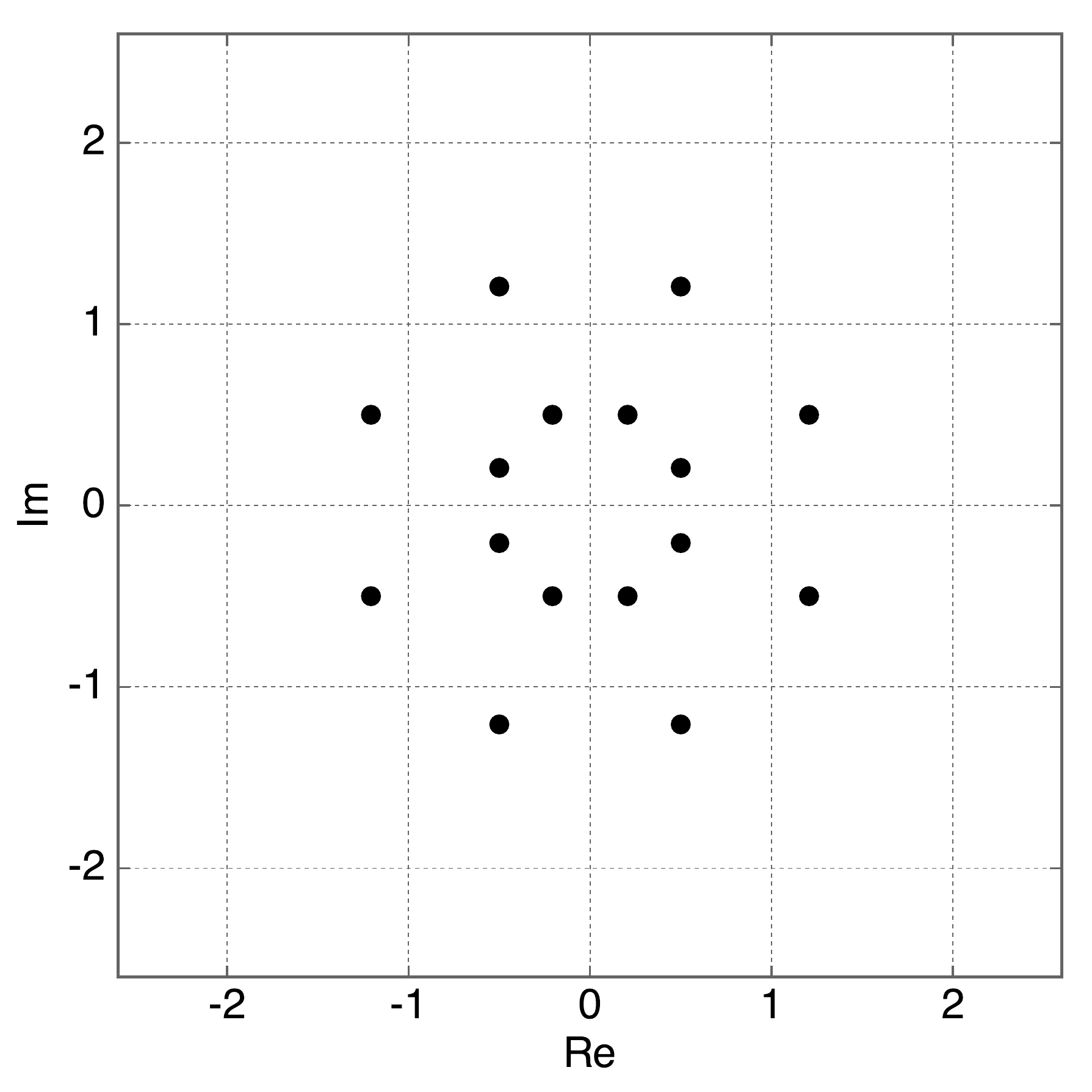}
				\label{const_FB11}
			}
			\subfigure[FB = 15]{%
				\includegraphics[width=0.2\hsize]{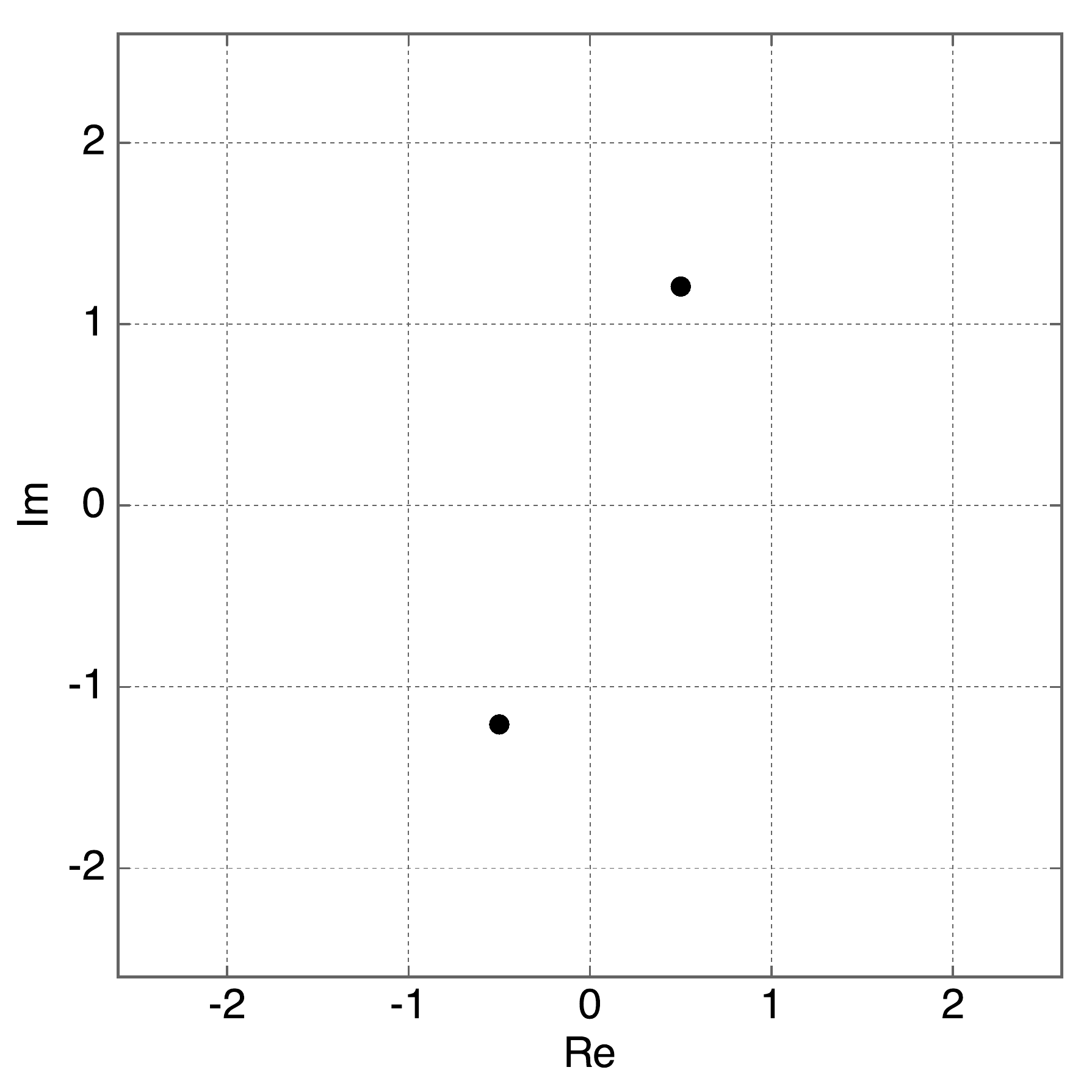}
				\label{const_FB15}
			}
			\caption{Filtered signal constellations of four filters for $\mathcal{\tilde{C}}(2,2)$.}
			\label{mapspace}
		\end{center}
	\end{figure*}

	The decoding performance of RASC depends on
	filter $g_1$ because the filter generates different signal
	constellations, as shown in Fig. \mbox{\ref{mapspace}}. These figures
	show various constellations with $g_1$
	for $L=2$ and $N_{bv}=2$.
	It appears from Figs. \mbox{\ref{const}}-\mbox{\subref{const_L2N2}} and
	\mbox{\ref{mapspace}}-\mbox{\subref{const_FB11}} that the constellation
	generated by $\text{FB}=11$ is same as $\tilde{C}(2,2)$. However, from Figs. \mbox{\ref{mapspace}}-\mbox{\subref{const_FB3}},
	\mbox{\ref{mapspace}}-\mbox{\subref{const_FB5}}, and \mbox{\ref{mapspace}}-\mbox{\subref{const_FB15}},
	the generated constellations corresponding to $\text{FB}=3$, $\text{FB}=5$ and $\text{FB}=15$ become a subset of Fig.~\mbox{\ref{const}}-\mbox{\subref{const_L2N2}}.
	If the constellation is a subset of $\tilde{C}(2,2)$,
	several filtered signal points degenerate into a few points.
	This implies that these filters lead to catastrophic codes because the degenerated signals cannot
	be correctly recovered from received signals. Therefore, bijective
	characteristics are a necessary condition when choosing filters.

	\subsubsection{Variable Rate with Variable Input Constraint}

	In turbo signal codes, the input signals must be restricted to $L^2$-QAM in order to obtain coding gains. The error-correcting capability of turbo signal codes depends on its component codes. When a set of input signals is identical to $C\left(L,N_{bv}\right)$, none of the individual component codes can provide any coding gain even if the number of filter taps is infinite, as demonstrated in \mbox{\cite{TSC}}. In contrast, RASC is based on a serial concatenation of a repetition code and a recursive code so that the coding gain can be obtained even if a set of input signals is identical to $C\left(L,N_{bv}\right)$ because the redundancy is introduced by the repetition code. From its definition, $\mathbb{Z}_L[j]$ is a subset of $C\left(L,N_{bv}\right)$ as the encoded signals of turbo signal codes.
	Hence, the input must be restricted to $L^2$-QAM, namely, $\mathbb{Z}_L[j]$.

	The transmission rate of turbo signal codes is defined by the information rate  of input $L^2$-QAM, namely $\log_2L^2$ \,bits per channel use (bpcu).
	Although $L$ must be increased to achieve a higher rate for turbo signal codes, large $L$ causes high decoding complexity because the number of states is given by $L^{2N_{bv}}$.
	However, especially when the input signals are chosen over $C\left(L,N_{bv}\right)$, our proposed RASC achieves an even higher transmission rate than turbo signal codes for the same $L$ because the cardinality of $\mathcal{\tilde{C}}\left(L,N_{bv}\right)$ is identical to that of $\mathcal{C}\left(L,N_{bv}\right)$ if and only if $N_{bv}=2$.
	Hence, the transmission rate of the non-systematic RASC is given by
	\begin{equation}
		R_c = \begin{cases}
			\displaystyle{\frac{\log_2L^2}{q}} & \text{if}\,\, \boldsymbol{s}\in
			\left(\mathbb{Z}_L\left[j\right]\right)^{N_s} \\ \\
			\displaystyle{\frac{\log_2L^4}{q}} & \text{if}\,\, \boldsymbol{s}\in
			\left(C\left(L,2\right)\right)^{N_s}.
		\end{cases}
	\end{equation}

	\subsection{Decoding Structure}
	\label{sec:dec}
	In this subsection, two decoding algorithms, namely the FFT-BP and EMS algorithm, are
	introduced for the RASC. These algorithms were originally proposed for
	decoding non-binary
	LDPCs defined over a Galois field with low computational complexity. In
	this paper, we modify these decoding algorithms for RASCs defined over $C\left(L,N_{bv}\right)$.

	\subsubsection{FFT-BP Algorithm}
	\label{sec:FFT-BP}
	The decoding complexity of the BCJR algorithm for the state-constrained signal
	codes is $O\left(L^{4N_{bv}}\right)$ because the number of
	states in the trellis diagram is given by $L^{2N_{bv}}$.
	Therefore, a low-complexity decoding algorithm is required, especially for large $L$ and $N_{bv}$. We hence employ the FFT-BP algorithm in RASC decoding. The complexity of this algorithm is reduced to  $O\left(L^{2N_{bv}}\log_2L^{2N_{bv}}\right)$.

	The exchanged message is defined as a vector of log-likelihood ratios
	(LLRs). The length of this vector is $L^{2N_{bv}}$, which is similar to that of non-binary LDPC codes.
	We first summarize the notations to describe the FFT-BP algorithm.

	{\bf Notations:}
	In an $m\times n$ parity matrix $\boldsymbol{H}_{m\times n}$, a set of non-zero elements of
	the row index at $n$ and a set of non-zero
	elements of the column index at $m$ are defined as
	$\mathcal{M}(n)\coloneqq \left\{n\mid H_{mn} \neq 0\right\}$ and
	$\mathcal{N}(m)\coloneqq \left\{m\mid H_{mn} \neq 0\right\}$,
	respectively. In addition, $\ell$ represents the number of decoding iterations.
	For ease of explanation, the proposed decoding algorithm is presented using a $6\times9$ small parity check matrix $\boldsymbol{H}_{6\times9}$ in (\mbox{\ref{eq:paritymatrix}}). Its corresponding bipartite graph is illustrated in Fig.~5.
	In this figure, the white circles at the top of the graph represent variable nodes of the information signal, The white square boxes in the middle represent check nodes, and the black circles at the bottom represent the variable nodes of parity signals.
	\begin{equation}
		\boldsymbol{H}_{6\times 9} = \left[\begin{array}[tb]{ccccccccc}
			1 & 0& 0& 1& 0& 0& 0& 0& 0\\
			0 & 1& 0& g_1& 1& 0& 0& 0& 0\\
			0 & 0& 1& 0& g_1& 1& 0& 0& 0\\
			1 & 0& 0& 0& 0& g_1& 1& 0& 0\\
			0 & 0& 1& 0& 0& 0& g_1& 1& 0\\
			0 & 1& 0& 0& 0& 0& 0& g_1& 1\\
		\end{array}\right]
		\label{eq:paritymatrix}
	\end{equation}

	\begin{figure}[tb]
		\begin{center}
			\includegraphics[width=\hsize]{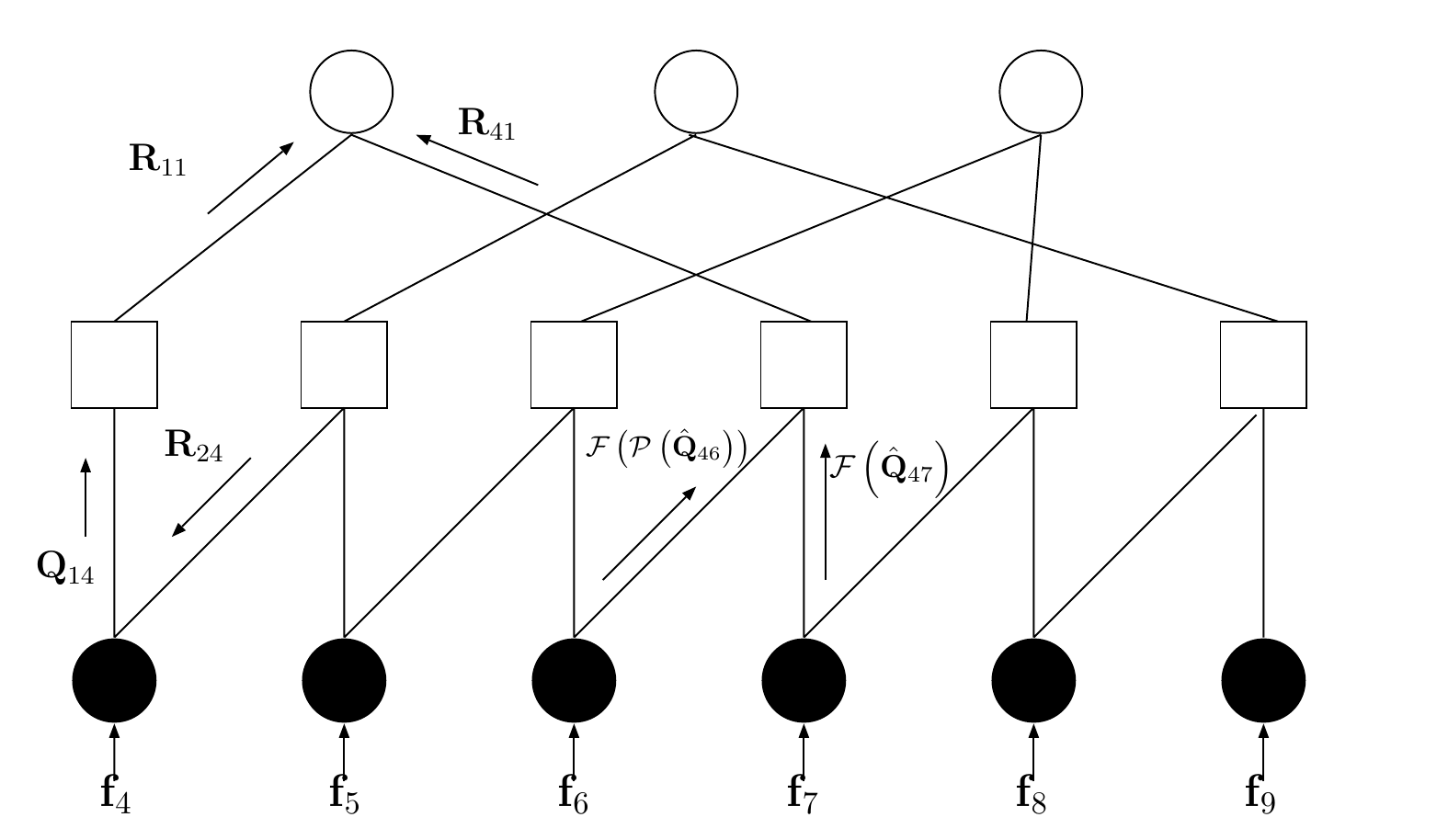}
			\caption{Bipartite graph of $\boldsymbol{H}_{6\times9}$ in (\ref{eq:paritymatrix})}
			\label{bipartite}
		\end{center}
	\end{figure}

	The message vectors of variable and check nodes at $\boldsymbol{H}_{mn}$
	are denoted by $\boldsymbol{Q}_{mn}^{\ell}$ and $\boldsymbol{R}_{mn}^{\ell}$,
	respectively. The channel LLR vector is also denoted by
	$\boldsymbol{f}_{n}$. As an example, the message vector of variable node  $\boldsymbol{Q}_{mn}^{\ell}$ is given by
	\begin{align}
		\boldsymbol{Q}_{mn}^{\ell} =
		\left[Q_{mn,a_0}^{\ell},\cdots,Q_{mn,a_{L^{2N_{bv}}-1}}^{\ell}\right]^T,
	\end{align}
	where
	\begin{align}
		Q_{mn,a_i}^{\ell} = {\displaystyle \log\frac{\Pr(X=a_i)}{\Pr(X=a_0)}}.
	\end{align}
	$P(X=a_i)$ is a probability that random variable $X$ takes on the value $a_i \in C\left(L,N_{bv}\right)$ for $i\in
	\left\{0,1,\cdots,L^{2N_{bv}}-1\right\}$. In addition, $Q_{mn,a_0}^{\ell}=0$ and
	$Q_{mn,a_i}^{\ell}\in \mathbb{R}$. Similarly, $\boldsymbol{R}_{mn}^{\ell}$ and
	$\boldsymbol{f}_n$ are expressed in the same manner.
	The four steps of the FFT-BP algorithm are described below.

	{\bf Initialization:}
	The initial values of all elements of $\boldsymbol{Q}_{mn}^{0}$ and
	$\boldsymbol{R}_{mn}^{0}$, and $\boldsymbol{f}_n$ are given by
	\begin{eqnarray}
		f_{n,a_i} &=& \log \frac{P\left(y_n|X=a_i\right)}{P(y_n|X=a_0)}, \\
		Q_{mn,a_i}^0 &=& 0, \\
		R_{mn,a_i}^0 &=& 0,
	\end{eqnarray}
	where $P\left(y_n|X=a_i\right)=(1/\sqrt{2\pi \sigma^2})\exp{\left((y_n - a_i)^2/2\sigma^2\right)}$.
	Note that $f_{n,a_i}=0, \forall i$, for $n\leq N_s$ in this paper because non-systematic RASC
	is considered. Thus, every LLR of the variable nodes corresponding to information signals is 0.

	{\bf Variable Node Updates:}
	The elements of variable node messages $Q_{mn,a_i}^{\ell}$ are updated via the following rule:
	\begin{eqnarray}
		Q_{mn,a_i}^{\ell}&=&f_{n,a_i} + \sum_{j\in{\cal M}\backslash m}R_{jn,a_i}^{\ell-1}-
		\alpha_{mn}, \nonumber \\
		\alpha_{mn} &=& \max_{a_i}Q_{mn,a_i}^{\ell},
		\label{eq:VN_obs}
	\end{eqnarray}
	where $\alpha_{mn}$ is a term for normalization that ensures numerical stability during the iterations.
	In non-systematic RASC, the variable nodes at $n \leq N_s$ are hidden nodes because the information signals are not transmitted.
	If the signals of hidden nodes are chosen over $s_i \in
	\mathbb{Z}_L[j]$, the message is updated as
	\begin{equation}
		Q_{mn,a_i}^{\ell}=\begin{cases}
			f_{n}+\sum_{j\in{\cal M}\backslash m}R_{jn,a_i}^{\ell-1}-\alpha_{mn} &
			a_i \in \mathbb{Z}_L[j]\\
			-\infty & \text{otherwise}.
		\end{cases}
		\label{eq:VN_hid}
	\end{equation}
	This procedure enables the decoder to ignore impossible candidates because the
	input signals are defined over $\mathbb{Z}_L[j]$. The main
	difference between the modified FFT-BP algorithm and the conventional one for non-binary LDPC codes is (\mbox{\ref{eq:VN_hid}})
	because the number of points in the input signal constellation can be less than that in the
	output signal constellation. If the input constellation is defined as
	$C\left(L,N_{bv}\right)$, RASC can be decoded
	by the same FFT-BP algorithm for a non-binary LDPC over GF($L^{2N_{bv}}$).

	{\bf Check Node Updates:}
	The update for both hidden and observation nodes is given by
	\begin{equation}
		\boldsymbol{R}_{mn}^{\ell} = {\cal P}^{-1}\left({\cal
			F}^{-1}\left(\sum_{j\in {\cal N}\backslash n}{\cal
			F}\left({\cal
			P}\left(\boldsymbol{Q}_{mj}^{\ell-1}\right)\right)\right)\right),
		\label{eq:CN_prc}
	\end{equation}
	where ${\cal P}$ indicates a permutation function, which depends on the filter coefficient $g_1$, ${\cal F}$ represents the multidimensional FFT
	function, and, ${\cal P}^{-1}$ and ${\cal F}^{-1}$ are inverse
	functions corresponding to ${\cal P}$ and ${\cal F}$,
	respectively. When $L=2$, ${\cal F}$ can be computed by the
	Walsh--Hadamard transform, just as in the decoding of non-binary LDPC codes \mbox{\cite{EMS}}.
	In addition, as described in \cite{FFT_Abelian}, there exists a Fourier transform if
	the code is constrained over Abelian groups. Therefore, if $L$ is equal to a power of two, RASC can be decoded via the FFT-BP algorithm because the codewords are constrained by $C(L,N_{bv})$. This also satisfies the property of commutative additive groups.
	Note that, in the log-domain FFT-BP algorithm, every LLR is split into the amplitude and sign for efficient and stable calculation, but details are omitted here for brevity. For more details, please refer to \cite{byers2004fourier}.

	{\bf Tentative Decision:}
	The information signal $s'_n$ at $n=1,2,\cdots,N_s$ is estimated as $a_i\in
	\mathbb{Z}_L[j]$ after ${\ell_{max}}$ iterations by the following criterion:
	\begin{equation}
		s'_n = \argmax_{a_i}f_{n,a_i}+\sum_{j\in{\cal M}}R_{jn,a_i}^{\ell_{max}}.
	\end{equation}

	To better elucidate the FFT-BP algorithm, we present the update calculations for $\mathbf{H}_{6\times 9}$. The corresponding Tanner graph is depicted in Fig. \mbox{\ref{bipartite}}.

	The variable node message from the first parity signal (the leftmost black circle) to the first check node (the leftmost white box) is given by
	\begin{eqnarray}
		Q_{14,a_i}^{\ell}&=&f_{4,a_i} + R_{24,a_i}^{\ell-1} - \alpha_{41}. \nonumber
	\end{eqnarray}
	In addition, the check node message of the first check node (the leftmost white box) to the first information node (the leftmost white circle) is given by
	\begin{eqnarray}
		\boldsymbol{R}_{11}^{\ell} = \boldsymbol{Q}_{14}^{\ell-1}.
		\label{eq:CN_ex1}
	\end{eqnarray}
	Similar to (\mbox{\ref{eq:CN_ex1}}), the check node message of the fourth check node (the fourth white box from the left) can be written as
	\begin{eqnarray}
		\boldsymbol{R}_{41}^{\ell} = {\cal F}^{-1}\left({\cal F}\left(\boldsymbol{Q}_{47}^{\ell-1}\right) +
		{\cal F}\left({\cal P}\left(\boldsymbol{Q}_{46}^{\ell-1}\right)\right) \right).
	\end{eqnarray}
	Finally, the tentative decision of the first information signal (the leftmost white circle) can be computed by
	\begin{equation}
		s'_1 = \argmax_{a_i}R_{11,a_i}^{\ell_{max}} + R_{41,a_i}^{\ell_{max}}.
	\end{equation}
	The other information signals can be also recovered in the same manner.

	\subsubsection{Modified EMS Algorithm}

	The FFT-BP algorithm can reduce the decoding complexity to
	$O\left(L^{2N_{bv}}\log_2L^{2N_{bv}}\right)$. However,
	the decoding complexity increases exponentially with respect to $L$ and $N_{{bv}}$.
	We hence introduce a modified EMS algorithm based on \cite{LEMS}
	to reduce this complexity.
	The decoding complexities of both the variable node and check node updates can
	be reduced by each of the elementary steps described below.
	Note that a sample code of the EMS decoder is available in \cite{EMS_github}.

	{\bf Message Truncation:}

	First, we truncate the message vectors to retain the largest
	$N_m$ LLRs. The truncated message vectors are denoted by $Q_{mn}[k]$ and $R_{mn}[k]$,
	$k=0,1,\cdots,N_m-1$. The values in these message vectors are sorted in
	decreasing order.
	Thus, $Q_{mn}[0] \geq Q_{mn}[1] \geq \cdots
	\geq Q_{mn}[N_m-1]$ must be satisfied (this is similarly true for $R_{mn}[k]$). The corresponding symbol for each $Q_{mn}[k]$ and
	$R_{mn}[k]$ must be stored along with the message vector.
	At every step, the modified EMS decoder chooses the $N_m$ largest messages. The decoder must store the correspondences between the symbols and stored messages, otherwise it cannot retrieve the original information using the partial messages. More specifically, the decoder does not know which signal corresponds to each $Q_{mn}[k]$ and $R_{mn}[k]$, so $\beta_{Q_{mn}[k]}$ and $\beta_{R_{mn}[k]}$ hold such information. For example, $\beta_{Q_{mn}[k]} \in C(L,N_{bv})$ stores the symbol corresponding to the message $Q_{mn}[k]$.
	However, the decoding complexity can be significantly reduced when $N_m\ll L^{2N_{bv}}$
	because only the truncated messages are exchanged.

	{\bf  Elementary Step of Variable Node Update:}

	We describe the computation of a variable node update in modified EMS in this subsection. We now assume
	$q=2$ for brevity. For $q>3$, the output of this elementary step is
	treated as an intermediate message and the step is processed recursively.
	The variable node update is illustrated in Fig. \mbox{\ref{EMS_steps}}-\mbox{\subref{EMS_V}}.
	The decoder first computes $2N_{m}$ message
		candidates from two sorted message vectors $\boldsymbol{R}_{m'n}$ and $\boldsymbol{R}_{m''n}$, which were computed in the previous iteration.
	These $2N_m$ candidates are temporarily stored in a $2N_m$-long message vector $\boldsymbol{T}$. Message vector $\boldsymbol{T}$ is sorted to find the $N_m$ most-reliable messages, and then the $k$th element of the sorted $\boldsymbol{T}$, $T[k]$ $(k=0,1,\cdots,N_m-1)$, is moved into the outgoing message $Q_{mn}[k]$. This procedure is expressed as follows:
	\begin{figure*}[tb]
		\begin{center}
			\subfigure[Variable node update.]{%
				\includegraphics[width=0.45\hsize]{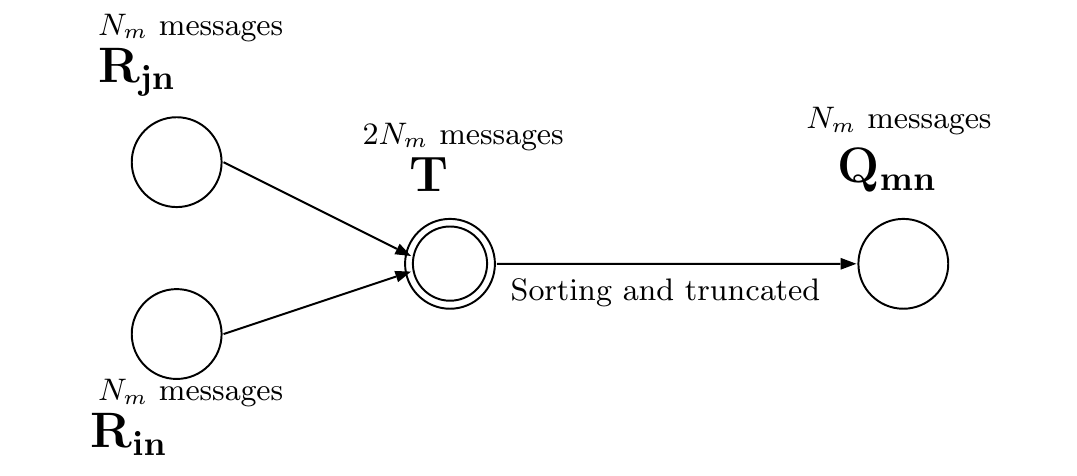}
				\label{EMS_V}
			}
			\subfigure[Check node update.]{%
				\includegraphics[width=0.45\hsize]{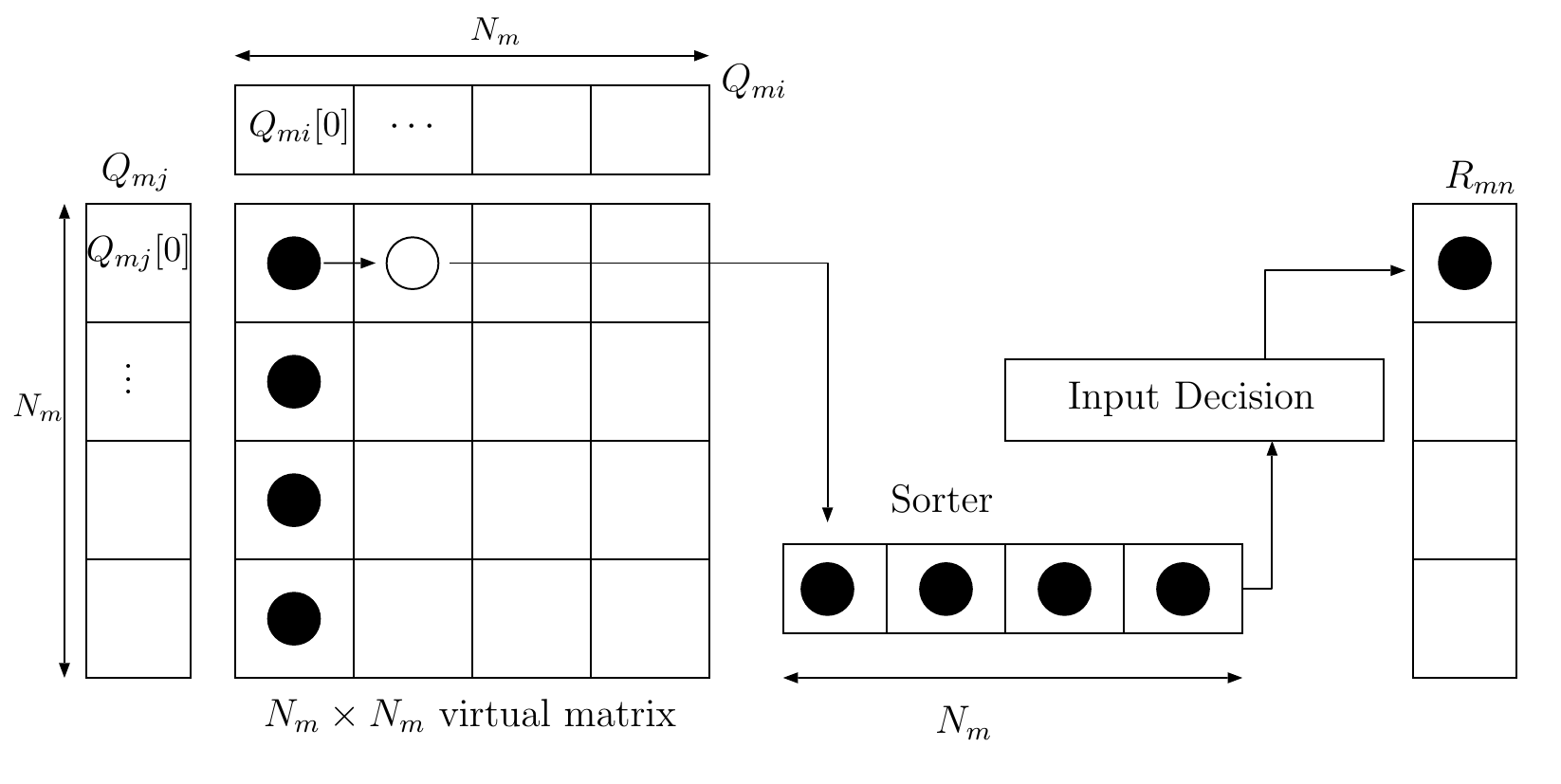}
				\label{EMS_C}
			}
			\caption{Elementary steps of the variable node and check node updates in the modified EMS algorithm}
			\label{EMS_steps}
		\end{center}
	\end{figure*}

	\begin{eqnarray}
		T[k] &=& R_{m'n}^{\ell-1}[k] + Y_{m''n}[k] \\
		T[k+N_m] &=& \gamma_{R_{m'n}}+R_{m''n}^{\ell-1}[k],
		\label{eq:EMS_varT}
	\end{eqnarray}
	where $k=0,1,\cdots,N_m-1$. Moreover, $Y_{m''n}[k]$ is given by
	\begin{eqnarray}
		Y_{m''n}[k] &=&\begin{cases}
			R_{m''n}^{\ell-1}[l] & \text{if} \,\,\, \beta_{R_{m''n}^{\ell-1}[l]} =
			\beta_{R_{m'n}^{\ell-1}[k]} \\
			\gamma_{R_{m''n}^{\ell-1}} & \text{if} \,\,\, \beta_{R_{m''n}^{\ell-1}[l]} \neq \beta_{R_{m'n}^{\ell-1}[k]},\,\,\, \forall l,k,
		\end{cases}
	\end{eqnarray}
	where $\gamma_{R_{m''n}}$ is a scalar value that compensates for the truncated
	$L^{2N_{bv}}-N_m$ LLRs and is given by
	\begin{equation}
		\gamma_{R_{m''n}} = R_{m''n}[N_m] - \log\left(L^{2N_m} - N_m\right) - \eta,
		\label{eq:gamma}
	\end{equation}
	where $\eta$ is an offset value optimized by density evolution \cite{LEMS}.
	The variable node message $Q_{mn}[k]$ consists of $N_m$ LLRs from
	$\boldsymbol{T}$. If the input signals are defined as $s_i\in
	\mathbb{Z}_l[j]$, then the hidden node messages are updated by

		\begin{equation}
			Q_{mn}[k] = \begin{cases}
				T[k] & \text{if} \,\,\, \beta_{T[k]} \in
				\mathbb{Z}_L[j] \\
				\gamma'_T & \text{if} \,\,\, \beta_{T[k]} \notin
				\mathbb{Z}_L[j] \text{ and } \\
				& \,\,\, \,\, \mathbb{Z}_L[j] \setminus \mathcal{S}\left(\beta_{\boldsymbol{Q}_{mn}}\right)\neq\{\} \\
				-\infty & \text{otherwise},
			\end{cases}
			\label{eq:VNU_Z}
		\end{equation}
		where $\mathcal{S}\left(\beta_{\boldsymbol{Q}_{mn}}\right)$ represents the set of all symbols of $\boldsymbol{Q}_{mn}$, $\mathbb{Z}_L[j] \setminus \mathcal{S}\left(\beta_{\boldsymbol{Q}_{mn}}\right)$ indicates the set $\mathbb{Z}_L[j]$ except for the set $\mathcal{S}\left(\beta_{\boldsymbol{Q}_{mn}}\right)$, $\{\}$ is defined as an empty set, and $\gamma'_T$ is given by
	\begin{equation}
		\gamma'_T = T[l_{z}] - \log\left(L^{2N_m} - l_z\right) - \eta,
	\end{equation}
	where $l_z$ represents the largest index of the LLR with the corresponding symbol $\beta_{T[l_z]}\in\mathbb{Z}[j]$.
	This algorithm slightly differs from the original EMS algorithm for non-binary LDPC codes because the decoder has \emph{a priori} knowledge of the input signals, which are constrained over $\mathbb{Z}_L[j]$. Therefore, the decoder replaces the LLR with negative infinity when the corresponding input signal is not in $\mathbb{Z}_L[j]$. The decoder also replaces the LLR with a constant value $\gamma'_T$ for the corresponding symbol of $\mathbb{Z}_L[j] \setminus \mathcal{S}\left(\beta_{\boldsymbol{Q}_{mn}}\right)$ from $-\infty$ with the signal not in $\mathbb{Z}_L[j]$.
	When the input signals are constrained in $C\left(L,N_{bv}\right)$,
	then the decoder naively updates the variable messages via (\ref{eq:EMS_varT}).


	{\bf Elementary Step of Check Node Update:}

	Similar to the variable node update, the output of this
	elementary step is stored as an intermediate message and recursively
	updates when the number of rows is greater than three.
	For simplicity, we assume two rows without loss of generality. Let $R_{mn}[k]$ be the outgoing message. Messages $Q_{mn'}[k']$ and $Q_{mn''}[k'']$
	are incoming messages for evaluating $R_{mn}[k]$.
	In the check node updates, the decoder computes the outgoing message vector
	from set $\mathcal{S}\left(\beta_{R_{mn}[k]}\right)$, defined as all possible symbol
	combinations that satisfy the parity equation $\beta_{Q_{mn'}[k']}+\beta_{Q_{mn''}[k'']}
	+\beta_{R_{mn}[k]}=0 \mod C\left(L,N_{bv}\right)$ as follows:
	\begin{eqnarray}
		R_{mn}[k] =
		\max_{\mathcal{S}\left(\beta_{R_{mn}[k]}\right)}\left(Q_{mn'}[k']+Q_{mn''}[k'']\right), \nonumber \\
		\,\,\,\,\,\,\,\,\,\,\,\,\,\, k,k',k''  \in \left\{0,1,\cdots,N_m-1\right\}. \nonumber
	\end{eqnarray}
	If the decoder naively searches for candidates in the set, the computational complexity would be $O(N_m^2)$.
	To reduce the computational complexity,
	we introduce an $N_m\times N_m$ virtual matrix (illustrated in Fig. \mbox{\ref{EMS_steps}}-\mbox{\subref{EMS_C}}).
	The element of this virtual matrix at $(k',k'')$ can be computed as $Q_{mn'}[k'] + Q_{mn''}[k'']$.
	The messages in the upper left of the virtual matrix are more reliable than those in
	the lower right because the incoming messages $Q_{mn'}[k']$ and $Q_{mn''}[k'']$ are
	sorted. Thus, the decoder can compute the outgoing message $R_{mn}[k]$ by finding $N_m$ messages from
	the upper left to the lower right of the virtual matrix.
	This algorithm consists of the following steps.
	\begin{enumerate}
		\item Give the LLR values of the first column and its corresponding symbol of the virtual matrix to the sorter.
		\item Determine the largest LLR value in the sorter.
		\item If the symbol associated with the largest output LLR value of the sorter exists in the output list, eliminate the largest LLR value and its corresponding symbol in the sorter and do not append them in the output list; otherwise, append the largest output LLR value and its corresponding symbol to the output list and eliminate them in the sorter.
\item In the virtual matrix, the LLR value and its corresponding symbol of the right neighbor of the largest LLR value in the sorter are moved to the sorter.
\item If the number of iterations is less than $2N_m$ and the length of the list is less than $N_m$, go to step 2. If the length of list does not reach $N_m$ while an iteration number is $2N_m$, the algorithm is stopped and the symbols and corresponding LLR of the sorter, except for the symbols already exist in the output list, are appended to the output list to reach the list length of $N_m$.
\end{enumerate}

The above check node update does not necessarily stop after $N_m$ steps because duplicate symbols can occur in step 3, and hence the
number of elements of the output list might not be $N_m$.  However, $2N_{m}$
steps are sufficient to ensure negligible degradation in the decoding
performance \cite{LEMS}.


\section{Noise Threshold Analysis and Filter Design}
\label{sec:Threshold}
\subsection{Monte Carlo Density Evolution}
Density evolution is a powerful tool for finding the noise threshold, which
is an important indicator of code performance.
In binary codes, the mathematical formulation of density evolution is easily obtained
because the belief-propagation (BP) messages are scalars \cite{DE_1, Quantize_DE}.
In the non-binary case, tracking the true BP message distribution
is impractically complex because the BP messages are vectors.
{\em Gaussian approximation} is a feasible way to track
the density of BP messages for non-binary LDPC codes because it can reduce
the number of parameters to only two, namely
the mean and the variance of the density  \cite{NB_DE_ARB}. However, the message
distribution of the check node diverges increasingly from the true distribution as the degree of the check node increases. Furthermore, Gaussian approximation can be used if and only if {\em channel symmetry}
and {\em permutation invariance} can be assumed \cite{NB-DE}. Channel
symmetry leads to an uncorrelated message distribution, so that the all-zero
codeword assumption may be valid. Permutation invariance means eliminating the
effect of the weight of the parity matrix, which is obtained by
a random weight coefficient.
In state-constrained signal codes, the BP messages also consist of multiple LLRs, and the decoding performance depends on the codewords because of the asymmetric property of non-uniform modulation. Therefore, we employ the technique of adding a random coset vector, as described in \cite{NB_DE_ARB}. The random-coset
vector is added at the end of the encoder. The random-coset
elements are randomly chosen and uniformly distributed over
$C\left(L,N_{bv}\right)$. Thus, the resulting output
codeword from the AWGN channel is symmetric. The proof of this
symmetric property resulting from the random-coset vector is omitted herein. (Please refer to \cite{NB_DE_ARB} for the proof.) Although the channel symmetry can be assumed,
permutation invariance cannot be assumed because the weight coefficients are
defined by the IIR filter coefficient.
MC-DE has been introduced as alternative approach for tracking the density
with multiple parameters \cite{Matteo_phd, DE_LDLC, SC-LDLC}.
One advantage of MC-DE is that the estimated noise threshold is more accurate than
the Gaussian approximation method because the analysis is non-parametric
\cite{Matteo_phd}. Thanks to the random-coset settings for the RASC, as described
above, we can straightforwardly introduce MC-DE to approximate noise
threshold $\sigma_{th}$.
We now briefly describe the MC-DE algorithm.
In this algorithm, four message pools $Q^I$, $Q^P$, $R^I$, and $R^P$ correspond to  the variable node messages of information signals, variable node messages of parity signals, check node messages of information signals, and check node messages of parity signals, respectively.
Every message pool has $N_{sam}$ messages. Message updates are performed by the modified FFT-BP algorithm described in Section~III-B.
When the check node message of the parity signal is calculated, the variable node message of the information signal and the variable node message
of the parity signal are uniformly picked up from message pools $Q^I$ and $Q^P$, respectively. Then, the updated check node message of
the parity signal is used as an input message to calculate the variable node message of the parity signal in $R^P$.
The other message updates can be also processed using the FFT-BP algorithm and the four message pools.

The parameters we set to calculate the noise threshold of RASC are as follows:
the number of the noise threshold calculations $R_{max}=10$, number of message
samples $N_{sam}=5,000$, maximum number of
iterations $\ell_{max}=100$, decoding error threshold
$P_{th}=10^{-4}$, which specifies that decoding could be regarded as
successful when all symbols are correctly decoded,
and noise threshold precision value $\epsilon_{\sigma}=10^{-5}$,
which is used to terminate the bisection method. If the interval of
searched noise variance is less than $\epsilon_{\sigma}$, the bisection search
algorithm is stopped.
In the subsequent section, we present the results of the optimum
parameters and the corresponding noise threshold.

\subsection{Searching for the Best Filter}
\label{sec:SBF}
As described in Section \ref{sec:intro}, an efficient search algorithm for the
filter coefficient of state-constrained signal codes has not been reported \cite{TSC}.
As shown in Fig. \mbox{\ref{mapspace}}, several filters that are not
bijective generate undecodable codewords. Therefore, the proposed search algorithm computes
the noise thresholds only for bijective filters so that the number of
candidate filters is less than $L^{2N_{bv}}$.
%

To clarify the relationships among filters,
several results for the best four filters with the parameters
found by MC-DE are shown in Tables~\mbox{\ref{tb:Fil_L2_N3}} and~\mbox{\ref{tb:Fil_L3_N2}}.
Several filters exhibit same or almost same noise thresholds. For instance, in Table~\mbox{\ref{tb:Fil_L2_N3}}, $\text{FB}=28$ and $\text{FB}=44$ have the same noise threshold (0.79\,dB). Similarly, $\text{FB}=52$ and $\text{FB}=56$ have the same noise threshold (0.80\,dB). The same relationships can be found in Table \mbox{\ref{tb:Fil_L3_N2}}. These filters are affine transformations of each other, e.g., rotated by 90 degrees, extensions, or reductions. This property can be exploited so as to further reduce the number of candidate filters.


\begin{table}[htbp]
				\caption{Best four filters for $L=2,N_{bv}=3$, and $q=2$.}
				\label{tb:Fil_L2_N3}
				\begin{center}
				\begin{tabular}{ccc}
					FB & Taps & Threshold\,[dB] \\ \hline
					28 & (0,0),(1,1),(1,0) & 0.79 \\
					44 & (0,0),(1,1),(0,1) & 0.79 \\
					52 & (0,0),(1,0),(1,1) & 0.80 \\
					56 & (0,0),(0,1),(1,1) & 0.80 \\ \hline
				\end{tabular}
			\end{center}
\end{table}

\begin{table}[htbp]
				\caption{Best four filters for $L=3,N_{bv}=2$, and $q=2$.}
				\label{tb:Fil_L3_N2}
				\begin{center}
				\begin{tabular}[tb]{ccc}
					FB & Taps & Threshold\,[dB] \\ \hline
					45 & (0,0),(2,1) & 3.85 \\
					63 & (0,0),(1,2) & 3.85 \\
					36 & (0,0),(1,1) & 3.86 \\
					72 & (0,0),(2,2) & 3.86 \\\hline
				\end{tabular}
			\end{center}
\end{table}

\subsection{Noise Thresholds for various $N_{bv}$}
\begin{figure}[t]
\begin{center}
	\includegraphics[width=\hsize]{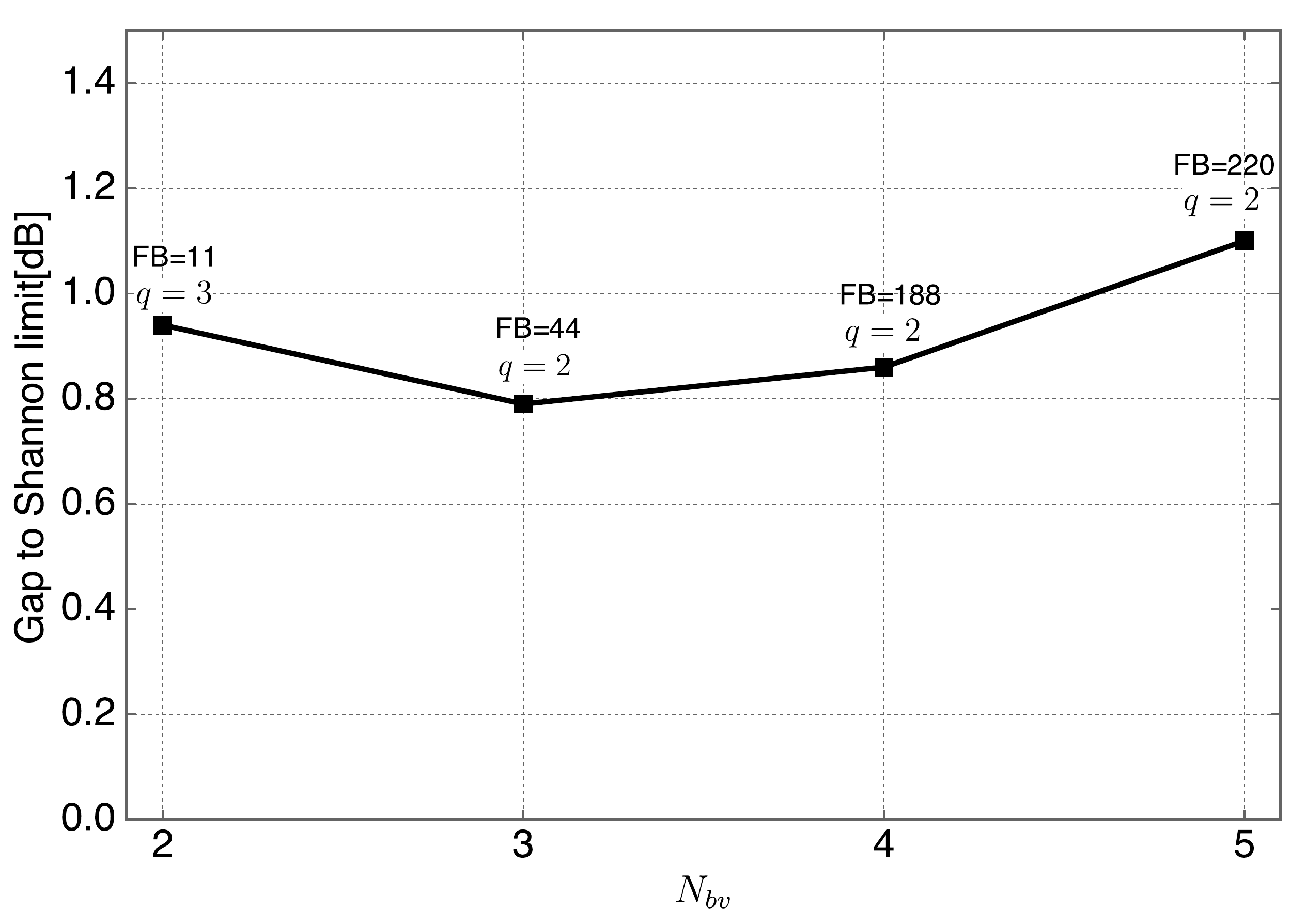}
	\caption{Noise thresholds of various $N_{bv}$ with $L=2$.}
	\label{Th_basis}
\end{center}
\end{figure}

Figure \ref{Th_basis} shows the difference between the noise threshold of the RASC and the Shannon limit.
In this paper, the Shannon limit is calculated using the channel capacity with a Gaussian input distribution.
This figure clarifies the impact of $N_{bv}$ on the noise threshold. The
indices of the best filters for each $N_{bv}$ are shown in this figure because
different values of $N_{bv}$ have different optimum filters.
Note that some affine-transformed filters have the same noise threshold, as
described above.
Thus, the filters must be determined for every target $N_{bv}$.
Furthermore, the number of repetitions $q$ must be
optimized for each $N_{bv}$. Based on our results, the
optimal value of $q$ is three for $N_{bv}=2$, whereas that for $N_{bv} \geq 3$ is two.
In non-binary LDPC, the optimum column weight is two
for a large alphabet size, e.g., greater than or equal to GF(64) \cite{davey2000error}.

Unfortunately, the noise threshold does not improve monotonically as $N_{bv}$
increases.
We believe that the reason for this is the dense constellation of output signals.
As shown in Figs. \ref{const}-\subref{const_L2N2},
\ref{const}-\subref{const_L2N3}, and \ref{const}-\subref{const_L2N4},
the distance between the signal points decreases as $N_{bv}$
increases. Thus, increasing $N_{bv}$ improves coding gain but results in a short Euclidean distance for the constellation.

\subsection{Noise Thresholds for Different Numbers of Repetitions}
\begin{figure}[tb]
\begin{center}
	\includegraphics[width=\hsize]{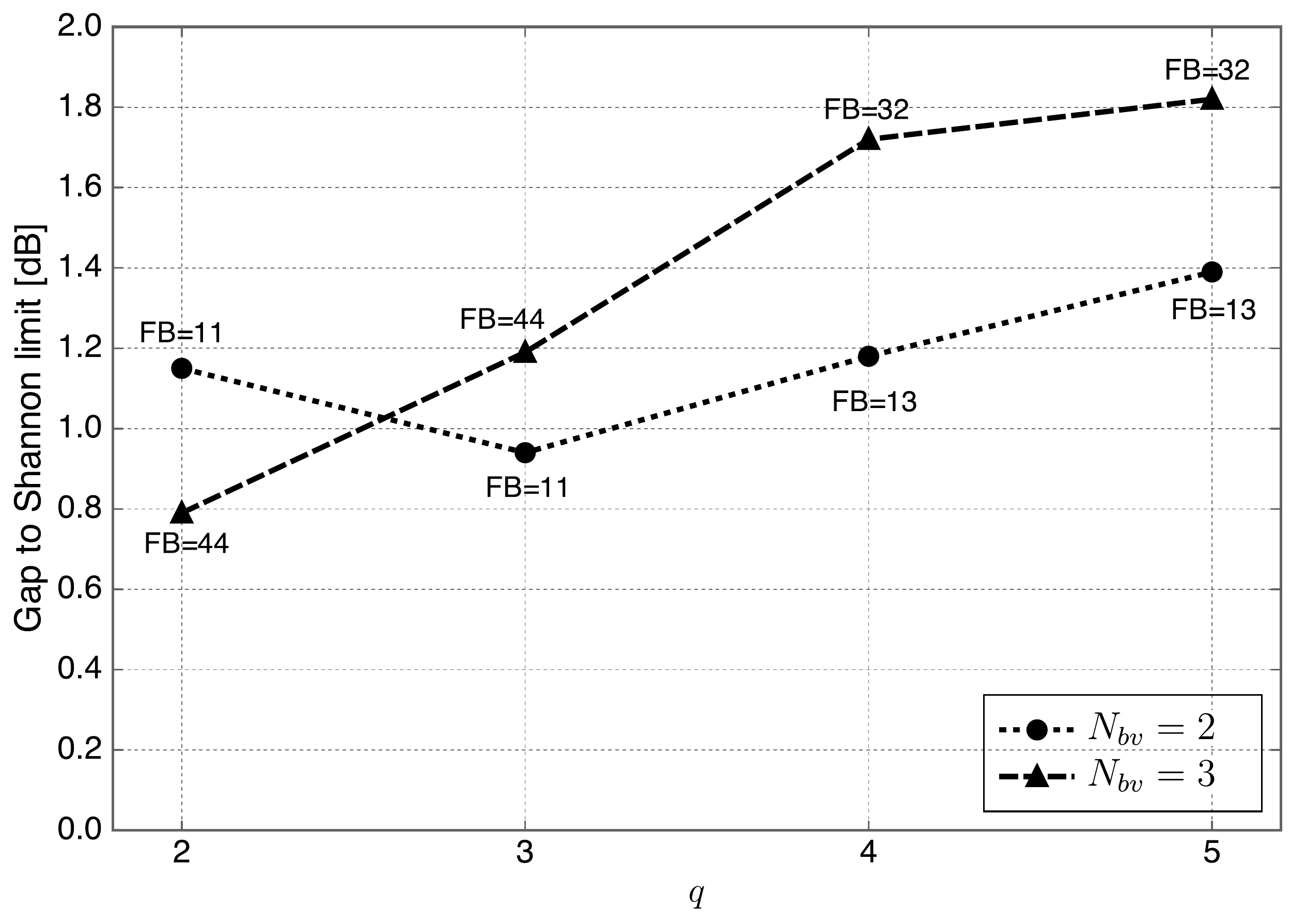}
	\caption{Noise thresholds of various $q$ for $L=2$ with $N_{bv}=2$ and $N_{bv}=3$.}
	\label{Th_column}
\end{center}
\end{figure}

The effect of the number of repetitions $q$ is shown in Fig. \ref{Th_column}.
Similar to the previous results, the optimum filter varies depending on $q$.
The difference between the noise threshold of the RASC and the Shannon limit
for $N_{bv}=3$ and $\text{FB}=44$ is the smallest for all parameter values at $q=2$ (= 0.79\,dB).
However, for $q \geq 3$, the noise thresholds for $N_{bv}=3$ are
inferior to those for $N_{bv}=2$. This conclusion can also be found in the
literature on non-binary LDPC codes, where, for a column weight greater than
three, the performance of the codes over higher-order fields is worse
than that of the codes over lower-order fields \cite{davey2000error}.

\subsection{Noise Thresholds for Different Input Constellation Sizes}
\begin{figure}[tb]
\begin{center}
	\includegraphics[width=\hsize]{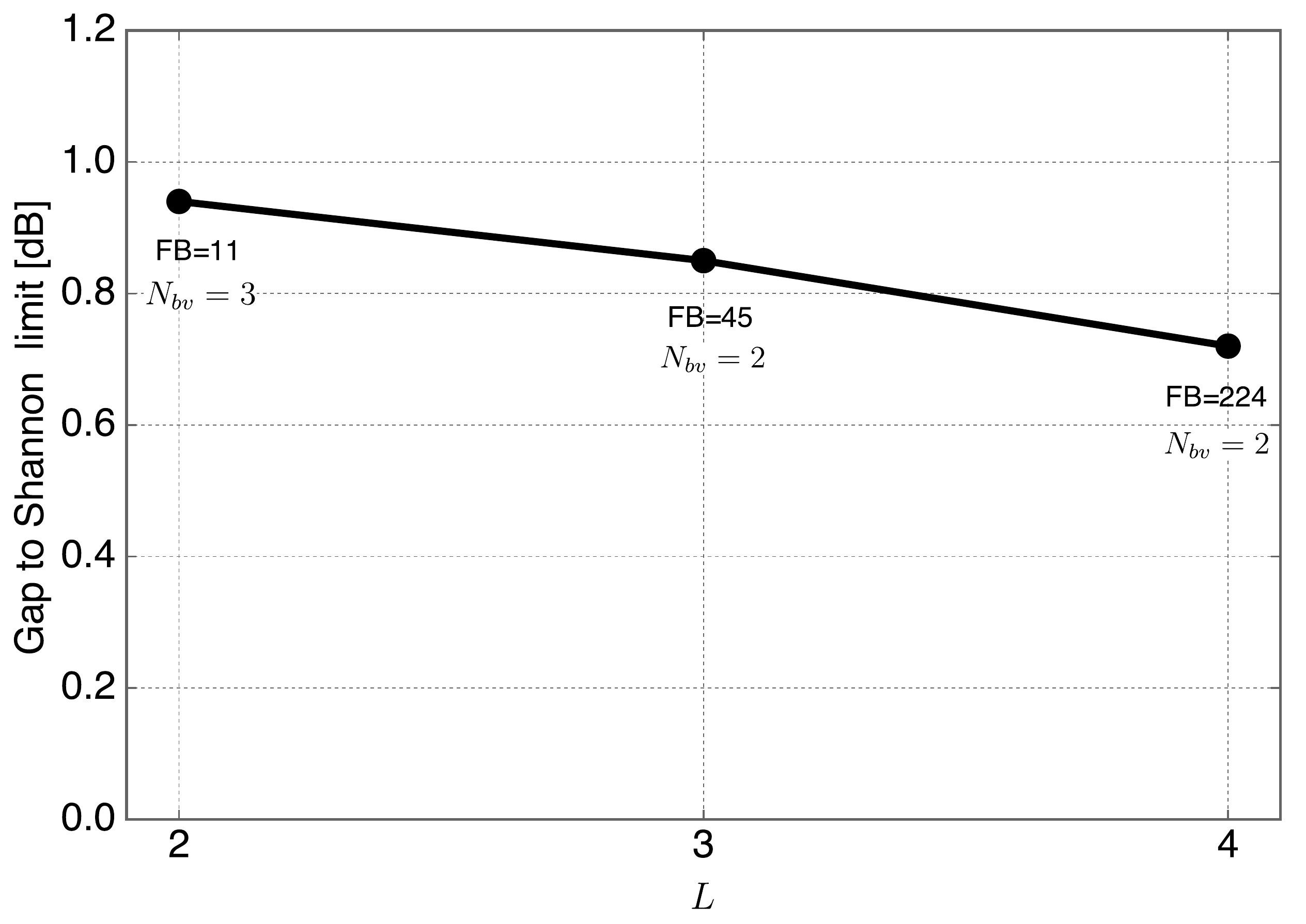}
	\caption{Noise thresholds for different input constellation sizes $L$. Here, $q=3$ is used for $L=2$ and $q=2$ is used for $L=3,4$.}
	\label{Th_L}
\end{center}
\end{figure}

The impact of the input constellation size $L$ is shown in Fig. \ref{Th_L}. Because the computational complexity required for $L \geq 3$
with $N_{bv} \geq 3$ is enormous, only the case of $N_{bv}=2$ is illustrated.
In this figure, taking into account the discussion about the optimum $q$ in the previous subsection, $q=3$ is used for $L=2$, and $q=2$ is used for $L=3,4$.
In contrast to the impact of $N_{bv}$, the difference between the noise threshold of the RASC and the Shannon capacity decreases as $L$ increases. The reason
for this behavior is that the
distances between the output signals are greater than for fixed $L$ because the signal constellation space is expanded by $L$. Hence, the RASC appears
to have the excellent property whereby increasing the transmission rate improves the noise threshold.

\subsection{Noise Threshold for Different Input Constraints}
\begin{figure}[tb]
\begin{center}
	\includegraphics[width=\hsize]{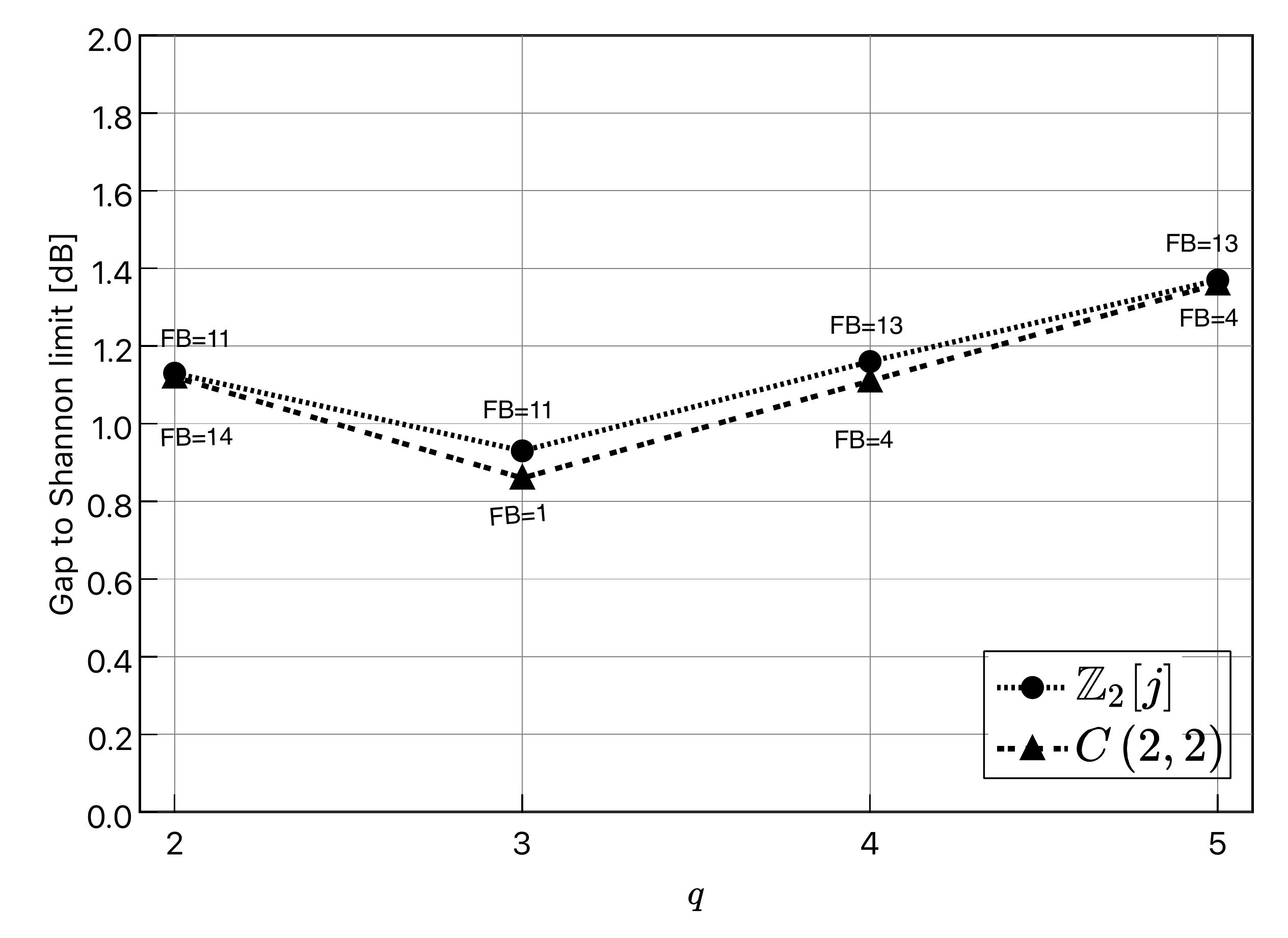}
	\caption{Noise thresholds for different input constraints for $L=2$ and $N_{bv}=2$}
	\label{Th_const}
\end{center}
\end{figure}

Finally, the effect of the input signal constraint is shown in
Fig. \ref{Th_const}. In this case, the optimum filters for the same $q$
and $N_{bv}$ differ depending on the constraint. Interestingly,
$\text{FB}=1$,  which is an identity mapping function, is the best filter for
$q=3$ and $L=N_{bv}=2$. We believe that the
reason for this is that when the input signals are chosen over $
C\left(L,N_{bv}\right)$, the transition of the state
due to the summation of signals is chosen over
$C\left(L,N_{bv}\right)$.
Then, the codeword distances increase, even if
the signals are mapped into the identical signal points by $\text{FB}=1$.

\section{Numerical Results}
\label{sec:Results}
In this section, the finite length performances of the RASC are shown.
We assume that the interleaver of the encoder is a random interleaver, the number of iterations of FFT-BP algorithm is 100, and the filters are chosen as described in the previous section.

\subsection{Performance of the RASC with FFT-BP}
\begin{figure}[tb]
\begin{center}
	\includegraphics[width=\hsize]{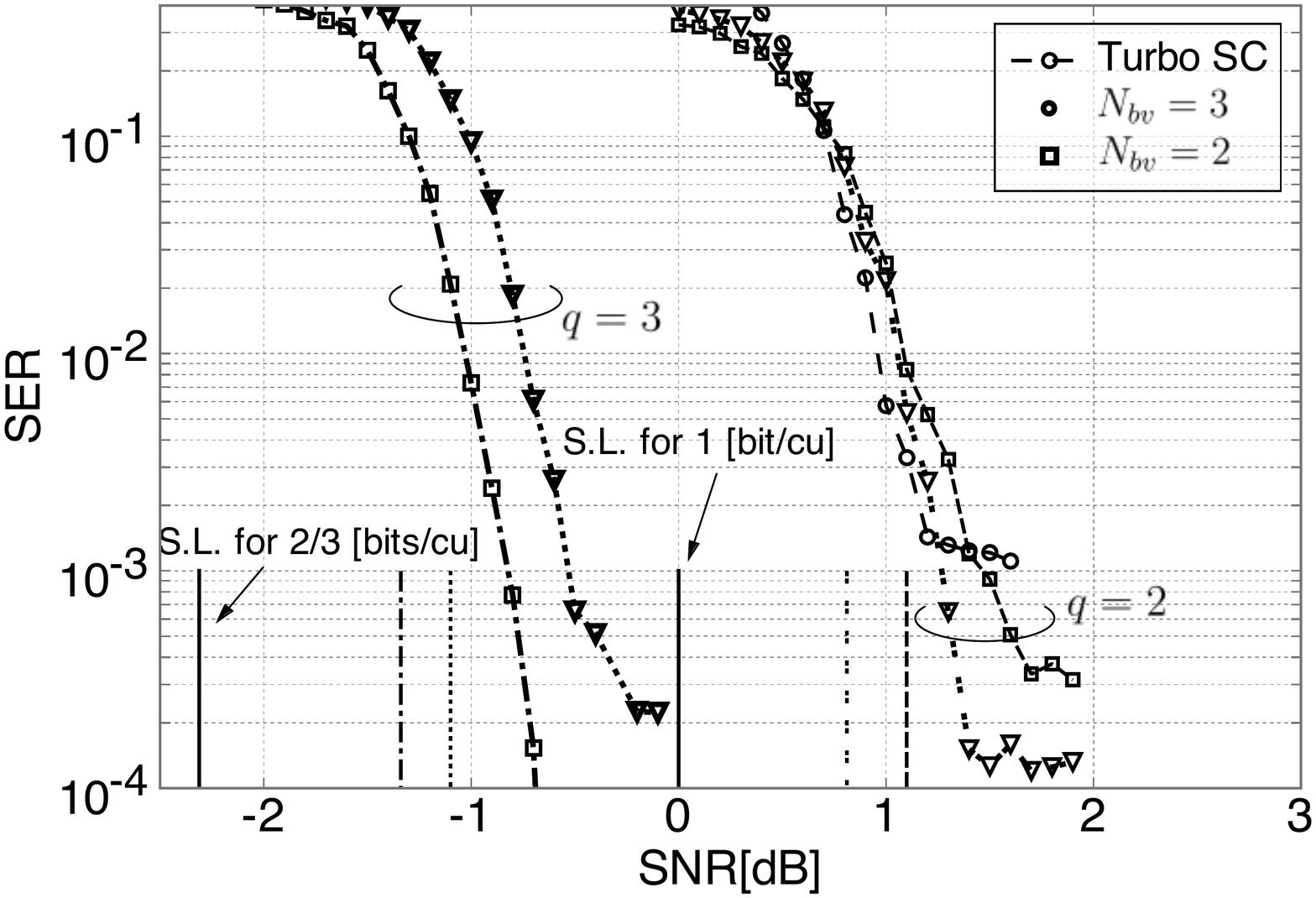}
	\caption{Performances of the optimized filter for $L=2$ and
		$N_{bv}=2,3$. The filter index is $\text{FB}=11$ for $N_{bv}=2$
		and $\text{FB}=44$ for $N_{bv}=3$. The solid lines indicate Shannon limits, and the other line styles indicate the noise thresholds corresponding to the same kinds of SER curves.}
	\label{SER_BP}
\end{center}
\end{figure}

Figure \ref{SER_BP} shows the SER of finite-codeword-length RASC with
FFT-BP decoding.
The input signals are 4-QAM ($L=2$) with length $N_s=1\,000$ and the
filters are chosen as $\text{FB}=11$ and
$\text{FB}=44$ for $N_{bv}=2$ and $N_{bv}=3$,
respectively.
As shown in the figure, the relationship
between $N_{bv}$ and $q$ is the same as that determined by a previous
noise threshold analysis. Namely, for $N_{bv}=2$, the performance
for $q=3$ is superior to that for $q=2$ in both the waterfall and error
floor regions. In contrast, for $N_{bv}\geq 3$, the noise threshold of $q=3$ is inferior
to that of $q=2$. Interestingly, the performance for the error
floor region depends on the filter but not the number of states.
Therefore, the best filter can only be chosen if we find the best noise threshold via MC-DE,
because the overall performance is defined by the
filter.

Next, we compare the performance of the RASC with that of the turbo signal codes where $N_s=1\,000$, $L=2$, and $N_{bv}=2$ for both codes. Due to a termination, the transmission rates of the RASC and the turbo
signal codes are $R_{RASC} = 2\times\frac{1,000}{2,000+1}=0.9995$\,bpcu. and $R_{turbo} =
2\times\frac{1,000}{2,000+8}=0.9960$\, bpcu, respectively.
For the turbo signal codes, we assume that the indices of the feedback and
feedforward filters are FB = 11 and FF = 81, respectively. This optimum
filter setting was reported in \cite{TSC}. The decoding algorithm is the BCJR algorithm, and
the number of iterations is 25. The order of computational complexity per
each decoding iteration of BCJR and FFT-BP are $O(L^{4N_{bv}})$ and
$O(L^{2N_{bv}}\log_2L^{2N_{bv}})$, respectively. Thus, for $L=2$ and
$N_{bv}=2$, the computational cost of BCJR is four times higher than
that of FFT-BP. Therefore, BCJR with 25 iterations and
FFT-BP with 100 iterations are equal in terms of decoding complexity.
As shown in Fig. \ref{SER_BP}, the performance of the RASC with $N_{bv}=2$ is close to that of turbo signal codes (within 0.2\,dB, which is
a negligible performance loss) in the waterfall region.
Furthermore, in the error floor region, the error probability of the RASC is
approximately $1/3$ better than that of the turbo signal codes. These
results indicate that RASCs can provide better performance than turbo signal
codes with only a one-tap feedback filter.

We finally mention the performance loss of our designed codes compared with the Shannon limit. The authors in \mbox{\cite{TSC}} showed that the mutual information for the constellations of $\tilde{C}(L,N_{bv})$ can almost achieve the channel capacity with Gaussian input distribution at a transmission rate equal to 1\,bpcu \mbox{\cite[Fig.~3]{TSC}}.
This implies that the shaping loss of RASC is negligible, and thus
the gap between the noise threshold of RASC and the Shannon limit mainly comes from the coding loss.

\subsection{Performance of the RASC with EMS}
\begin{figure}[tb]
\begin{center}
	\includegraphics[width=\hsize]{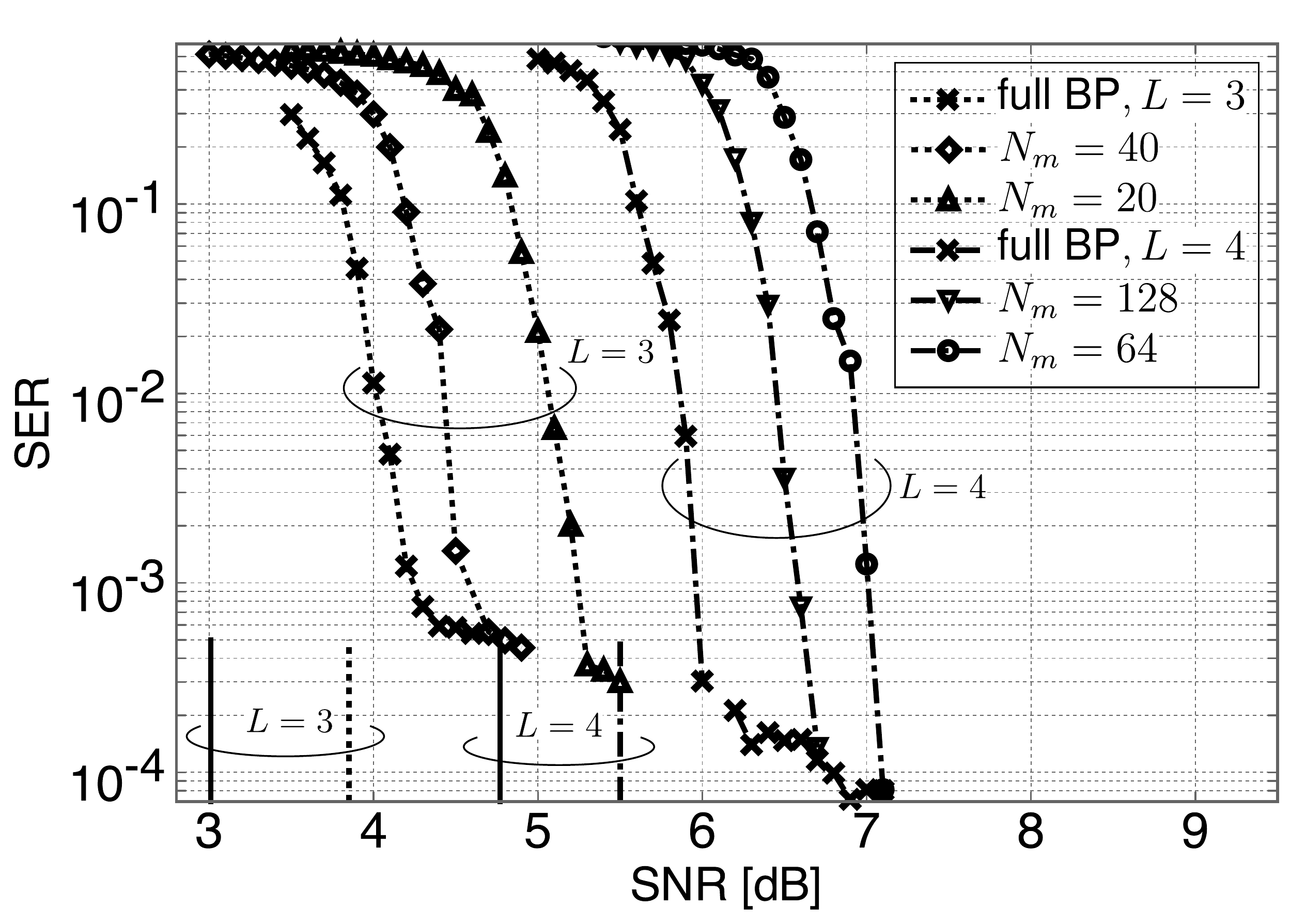}
	\caption{Performance with the EMS decoder for different values of $L$
		and $N_m$. For all curves, $N_{bv}$=2. The filter index is
		$\text{FB}=45$ for $L=3$ and $\text{FB}=224$ for $L=4$. $\eta=\{-3.0,
		-3.5, -4.7, -5.5 \}$ for
		$N_m=\{20, 40, 64, 128\}$, respectively. The solid lines indicate
		Shannon limits for each $L$. The dotted line and the chained line indicate the noise thresholds for $L=3$ and $L=4$, respectively.}
	\label{SER_EMS}
\end{center}
\end{figure}

Figure \ref{SER_EMS} shows the SER performance for $L=3$ and $L=4$
obtained using the proposed modified EMS decoder. We assume that $N_{bv}=2$, $q=2$, and $N_s=1\,000$.
In this figure, the dotted line and the chained line indicate the noise threshold for $L=3$ and $L=4$ with the optimized filters.
\emph{Full BP} indicates the naive sum-product
algorithm whereby all sets of the signals that satisfy the check equation are
searched and calculated for the message updates in check nodes. This is so that
the decoder does not use the FFT-BP algorithm because the FFT algorithm over
$C(L,N_{bv})$ with $L > 2$ cannot be represented in a multidimensional
FFT form as it can with $L=2$.
The number of output constellation points $L^{2N_{bv}}$ for $L=3$ and
$L=4$ are 81 and 256, respectively.
The decoding complexities of the BCJR, FFT-BP, and EMS algorithms are
$O(L^{4N_{bv}})$, $O(L^{2N_{bv}}\log_2L^{2N_{bv}})$, and
$O(N_m\log_2N_m)$, respectively. Therefore, the decoding complexity of
EMS can be reduced if $N_m$ is chosen to be much smaller than $L^{2N_{bv}}$. For
example, when $N_m=L^{2N_{bv}}/2$,
the decoding complexity of EMS becomes a quarter of that of the BCJR algorithm, and when $N_m=L^{2N_{bv}}/4$, the decoding complexity of the EMS becomes one-sixteenth
that of the BCJR algorithm and a quarter of that of FFT-BP algorithm.
Therefore, in this section, $N_m=40$ and $N_{m}=20$ are used for $L=3$
and, $N_m=128$ and $N_m=64$ are used for $L=4$.
Figure \ref{SER_EMS} indicates that the performance loss is about 0.5\,dB when $N_m=L^{2N_{bv}}/2$ and is about 1.0\,dB when $N_m=L^{2N_{bv}}/4$. Furthermore, our algorithm does not degrade the error floor performance because the curves with the EMS decoder decrease until reaching the error floor of a full BP. As a result, the EMS decoder can dramatically reduce the
decoding complexity without significant performance loss.

\section{Conclusion}
\label{sec:Conclusion}
In this paper, we proposed a new state-constrained signal code
based on RA codes. The proposed code has several
advantages compared with turbo signal codes, including a simpler encoder and decoder,
approximately the same performance in the waterfall region, and a slightly better
performance in the error floor region.
Furthermore, we found a design criterion for the filter coefficient
via MC-DE. We summarize the properties of
the noise threshold in terms of the following three parameters:
\begin{enumerate}
\item The parameter of the ring of formal power series $N_{bv}$
can improve the noise threshold, but an excessive value of $N_{bv}$ degrades it.
\item An optimum repetition number $q$ exists for a given
$N_{bv}$. In particular, for a large number of states, $L^{2N_{bv}}
\geq 64$, $q=2$ is optimum in terms of the noise threshold.
\item The number of states $L$ can improve both the transmission rate
and the noise threshold.
\end{enumerate}
The simulation results showed that the optimum filters for given
values of $N_{bv}$, $q$, and $L$ perform well in the waterfall and error floor regions
for short codeword lengths.
The simulation results also indicate that the performance loss due to the modified EMS decoder is less than 1\,dB as compared to the performance of the full BP
decoder, whereas the decoding complexity is reduced to less than 25\% of
that of the BCJR and FFT-BP algorithms.
Finally, we note that the proposed code can be easily generalized to irregular codes and other turbo-like codes such as accumulate-repeat-accumulate codes and braided
convolutional codes, to achieve more close to the Shannon limit. Furthermore, higher code rate design and insight of code properties such as shaping loss remain as the future work.


%

\appendices
\section{Necessary condition to satisfy the parity check equation in
turbo signal codes}
\label{sec:pr_turbo}
For simplicity, we consider an encoder that has two feedforward filters and one feedback filter
as follows:
\begin{equation}
x_t = f_1u_{t-1}+f_0\left(s_t+g_1u_{t-1}+b_t\right).
\label{eq:tsc_min}
\end{equation}
In this setting, the parity check matrix of (\ref{eq:tsc_min}) is given by
\begin{equation}
\left[\begin{array}[tb]{ccccccc|ccccccc}
	1 & 0& 0& \cdots& 0& 0& 0& 1& 0& 0& \cdots& 0 & 0& 0\\
	1 & 1& 0& \ddots& 0& 0& 0& 1& 1& 0& \ddots& 0 & 0& 0\\
	0 & 1& 1& \ddots& 0& 0& 0& 0& 1& 1& \ddots& 0 & 0& 0\\
	0 & 0& 0& \ddots& 1& 1& 0& 0& 0& 0& \ddots& 1 & 1& 0\\
	0 & 0& 0& \ddots& 0& 1& 1& 0& 0& 0& \ddots& 0& 1& 1\\
\end{array}\right],
\end{equation}
where the left-hand side of this matrix represents information
signals and the right-hand side represents parity signals.
From this matrix, the parity check equation at $t=1$ is given by
\begin{eqnarray}
s_0+s_1+x_0+x_1 &=& 0\mod C\left(L,N_{bv}\right).
\label{eq:parity}
\end{eqnarray}
From (\ref{eq:tsc_min}), the reformulation of (\ref{eq:parity}) is given by
\begin{equation}
(1+f_0+f_0g_1+f_1)s_0+\left(1+f_0\right)s_1 = 0\mod C\left(L,N_{bv}\right).
\end{equation}
Therefore, the filters for satisfying above parity check equation must be $f_0 = -1+j0 \mod C\left(L,N_{bv}\right)$ and $f_1 = g_1$.

For $L=2$, the coded signal at $t$ is given by
\begin{eqnarray}
x_t &=& f_1u_{t-1}+\left(s_t+f_1u_{t-1}+b_t\right) = s_t.
\label{eq:tsc_enc_signal}
\end{eqnarray}
Therefore, in turbo signal codes, there is no coding gain if
the signals are constrained by the parity check equation.
This is because the filter does not affect the input signal, as shown in (\ref{eq:tsc_enc_signal}).



\ifCLASSOPTIONcaptionsoff
\newpage
\fi



\bibliographystyle{IEEEtranTCOM}
\bibliography{myref}
\end{document}